\documentclass[11pt,letterpaper]{article}

\usepackage{jcappub}
\usepackage{times}
\usepackage{amsfonts}
\usepackage{latexsym}
\usepackage{dcolumn}
\usepackage{bm}
\usepackage{empheq}
\usepackage{bbold}

\usepackage[letterpaper]{geometry}
\usepackage{color}

\def\beq{\begin{equation}}
\def\eeq{\end{equation}}

\def\k{x_*}
\def\x{x_*}
\def\tmu{\tilde{\mu}}

\def\tmu{{\tilde \mu}}
\def\F{{\cal F}}
\def\m{\hat m}

\title{Influence of heavy modes on perturbations in multiple field inflation}

\author[a,b,c]{Xian Gao}%
\author[a]{David Langlois}%
\author[d,a]{Shuntaro Mizuno}%

\affiliation[a]{%
        \href{http://www.apc.univ-paris7.fr/APC_CS/en}{Astroparticule \& Cosmologie}, UMR 7164-CNRS, Universit\'{e} Denis Diderot-Paris 7,
        10 rue Alice Domon et L\'{e}onie Duquet, 75205 Paris, France}

\affiliation[b]{%
        \href{http://www.iap.fr/english/}{${\mathcal{G}}{\mathbb{R}}
\varepsilon{\mathbb{C}}{\mathcal{O}}$, Institut d'Astrophysique de Paris}, UMR 7095-CNRS, Universit\'{e} Pierre et Marie Curie-Paris 6, 98bis Boulevard Arago, 75014 Paris, France
        }%
        
\affiliation[c]{%
        \href{http://www.lpt.ens.fr/?lang=en}{Laboratoire de Physique Th\'{e}orique, \'{E}cole Normale Sup\'{e}rieure}, 24 rue Lhomond, 75231 Paris, France
        }%
        
\affiliation[d]{%
        \href{http://www.th.u-psud.fr}{Laboratoire de Physique Th\'{e}orique, Universit\'{e} Paris-Sud 11 et CNRS}, B\^{a}timent 210, 91405 Orsay Cedex, France
        }%

\emailAdd{xgao@apc.univ-paris7.fr}
\emailAdd{langlois@apc.univ-paris7.fr}
\emailAdd{shuntaro.mizuno@th.u-psud.fr}

\date{\today}


\abstract{
  We investigate linear cosmological perturbations in multiple field inflationary models where some of the directions are light while others are heavy (with respect to the Hubble parameter).   By integrating out  the massive degrees of freedom,
  we  determine the multi-dimensional effective  theory for the light degrees of freedom and give explicitly the propagation matrix that replaces the effective sound speed of the one-dimensional case.
   We then examine in detail the consequences of  a sudden turn along the inflationary trajectory, in particular the possible breakdown of the low energy effective theory in case the heavy modes are excited. Resorting to a new basis in field space, instead of the usual adiabatic/entropic basis, we study the evolution of the perturbations during the turn. 
In particular, we compute the  power spectrum and compare with  the  result obtained from the low energy effective theory.
}

\begin{document}
\maketitle%

\section{Introduction}
Although the inflation paradigm has  
been reinforced by the  accumulation of new cosmological data since its emergence more than  thirty years ago, 
 the nature of the inflaton field(s) remains elusive. In high energy physics models, it is generic to obtain a large number of scalar fields and inflation can be realized if, in this large field space, there exist  one or several directions along which the effective mass is small with respect to the Hubble parameter $H$. In addition to these light directions,  one 
 usually finds  other directions where the effective mass is of the order of $H$, or  much larger. Until recently, the standard approach was just to ignore these massive directions, since the perturbations generated along these directions are suppressed.

However, it was realized that the heavy modes can influence the evolution of the light modes if the background trajectory in field space is curved. In particular, it was shown  in the context of two-field models,  that the coupling between the light and heavy  degrees of freedom can be described  as a single-field system with an effective speed of sound that differs from the light speed when the inflationary trajectory undergoes a turn \cite{Tolley:2009fg}, even if the kinetic terms in the original two-field Lagrangian are perfectly canonical. This property was further investigated in \cite{Achucarro:2010jv,Achucarro:2010da}, as well as in \cite{Cremonini:2010ua,Achucarro:2012sm,Chen:2012ge,Pi:2012gf,Avgoustidis:2012yc,Achucarro:2012yr}. Moreover, when the turn of the trajectory is too sharp,  the single field effective  description breaks down, as studied  in \cite{Shiu:2011qw} and  \cite{Cespedes:2012hu}, because  the massive degrees of freedom  become excited during the turn (see also \cite{Jackson:2010cw,Jackson:2011qg,Jackson:2012fu} for recent works investigating the effects of heavy modes from another perspective).

The goal of the present work is to extend the above results  in several directions. Firstly, whereas previous studies usually work in the framework of  the traditional adiabatic-entropic basis~\cite{Gordon:2000hv,GNvT},  we advocate the use of a new basis for a better intuitive understanding of the behaviour of the background trajectory as well as of the perturbations. This is motivated by the consideration that the background velocity, i.e. the adiabatic direction, does not always coincide with the light direction, as emphasized in \cite{Shiu:2011qw}.
We thus prefer to use a basis where the distinction between light and heavy modes is manifest.
This basis, which we call the {\it mass basis}, corresponds to the eigenvectors of  the effective mass matrix which appears in the Lagrangian governing the linear perturbations of the system.  In many cases, this basis almost coincides with what we call the {\it potential basis}, which are the eigenvectors of the matrix whose components are the second derivatives of the potential with respect to the scalar fields, i.e. the Hessian of the potential.
Working in this mass basis, we obtain the effective action governing the dynamics of an arbitrary number of light (i.e. $m\ll H$) degrees of freedom in presence of an arbitrary number of heavy (i.e. $m\gg H$) degrees of freedom.  In this generalized framework, the effective speed of sound that was obtained previously in the case of a single light degree of freedom, becomes a matrix, whose eigenvalues correspond to different effective speeds of sound.

Secondly, we explore the regime where the low-energy effective theory breaks down. Indeed, our  multi-dimensional effective description is only valid as long as the inflationary trajectory remains relatively smooth.
In some models, one could imagine that the potential landscape is sufficiently hilly that  the inflationary trajectory has to undergo some sharp turns at some particular points. During these sharp turns, one must take into account explicitly the effects of the heavy degrees of freedom, as pointed out in \cite{Shiu:2011qw}.
In the present work, we explore in detail what happens during such a turn, considering a two-dimensional description in the two-dimensional plane in field space spanned by the inflationary velocities  just before and after the turn.  We also  assume that the trajectory is straight before and after the turn\footnote{This differs from a few works~\cite{Chen:2009we,Chen:2009zp,Gao:2009qy,Chen:2012ge,Pi:2012gf}, where the trajectory follows a circle with constant angular velocity (which we refer to as ``constant turn").}.
In order to develop an intuitive understanding of the consequences of  the turn on the perturbations, we build a simplified  analytical model of the turn,  which contains a parameter controlling the ``sharpness'' of the turn.
When the turn is smooth enough, the single-mode effective description is sufficient to compute the power spectrum. By contrast, for a sharp turn, it is necessary  to go beyond the effective single field description and we obtain an analytical expression for the contribution due to massive excitations. Both light and heavy contributions contain oscillations periodic in $k$.
We also solve numerically the equations of motion in a  concrete model, which we compare with our analytical description.

The outline of the paper is the following. In the next section, Section \ref{sec:eom}, we introduce our multi-field model and derive the equations of motion for the background and for the linear perturbations. Section \ref{sec:eff} is devoted to the derivation of the multi-dimensional effective theory when several light directions coexist. In section \ref{sec:turn}, we study the inflationary trajectory during a turn. Section \ref{sec:perts} deals with the evolution of the perturbations during this turn. We conclude in the final section and give some details about our calculations and approximations in the Appendix.

\section{Equations of motion for the background and the perturbations}{\label{sec:eom}}
For simplicity, we restrict ourselves to   multi-field models with {\it canonical} kinetic terms, although the extension to models with non-canonical kinetic terms (following e.g. \cite{LRPST}) should be straightforward.
We thus consider $N$ scalar fields $\phi^I$, governed by the action
    \begin{equation}
    \label{action}
        S=\int d^{4}x\sqrt{-g}\left(-\frac{1}{2}\delta_{IJ}\partial_{\mu}\phi^{I}\partial^{\mu}\phi^{J}-V\left(\phi^I\right)\right),\qquad I=1,2,\cdots,N,
    \end{equation}
where $g$ is the determinant of the spacetime metric $g_{\mu\nu}$ and $V(\phi^I)$ is the potential of the scalar fields.
We use  Einstein's implicit summation rule for the scalar field indices.

\subsection{Background equations of motion}
In a spatially homogeneous and isotropic spacetime, endowed  with the  metric
\beq
ds^2=g_{\mu\nu}dx^\mu dx^\nu=-dt^2+a^2(t)\delta_{ij}dx^idx^j,
\eeq
the evolution of  the scale factor $a(t)$ is governed by  the Friedmann equations
    \begin{equation}
H^{2}=\frac{1}{3}\left(\frac{1}{2}\delta_{IJ}\dot{\phi}^{I}\dot{\phi}^{J}+V\right),\qquad\qquad \dot{H}=-\frac{1}{2}\delta_{IJ}\dot{\phi}^{I}\dot{\phi}^{J}\equiv-H^{2}\epsilon,\label{eom_bg}
\end{equation}
where $H\equiv \dot a/a$ is the Hubble parameter and a dot denotes a derivative with respect to the cosmic time $t$.
We work in units such that $M_P\equiv (8\pi G)^{-1/2}=1$. The
equations of motion for the homogeneous scalar fields are
\begin{equation}
\ddot{\phi}^{I}+3H\dot{\phi}^{I}+\delta^{IJ}V_{,J}=0\,,\label{phi_eom_pb}
\end{equation}
where $V_{,J}\equiv \partial V/\partial \phi^J$.

As we will see later, it is often convenient to make a change of orthonormal basis in field space, so that
    \begin{equation}
        \phi^I = \phi^m e^I_m,\qquad \qquad m=1,\cdots,N, \label{change_basis}
    \end{equation}
    where the new basis vectors $e^I_m$ satisfy  the orthonormality condition
    \beq
    \label{orthonormality}
    \delta_{IJ}\, e^I_me^J_n=\delta_{mn},
    \eeq
    as well as  $\delta^{mn}e^I_me^J_n = \delta^{IJ}$.

In terms of    the new components $\phi^m$ of the inflationary trajectory,
the equations of motion (\ref{phi_eom_pb})  become
\begin{equation}
\label{eom_bck_gen}
        D_t\dot{\phi}_m +3H \dot{\phi}_m +V_{,m}=0,
    \end{equation}
where
\beq
\label{Z_mn_def}
D_t \dot{\phi}_m = \ddot{\phi}_m +Z_{mn}\dot{\phi}_n\,,\qquad
Z_{mn}:= \delta_{IJ} e^I_m \dot{e}^J_n\,.
\eeq
The coefficients $Z_{mn}$, which  are antisymmetric ($Z_{mn}=-Z_{nm}$)   as a consequence of the orthonormality condition (\ref{orthonormality}),  characterize the change rate of the new basis with respect to the former field space basis. For example, in a two-dimensional field space, the new basis is fully characterized by its angle $\theta$ with respect to the initial basis, and the coefficients $Z_{mn}$ are then directly related to $\dot\theta$.

\subsection{Linear perturbations}
Let us now turn to the linear perturbations of the $N$ scalar fields coupled to gravity.
For  multi-field models, it is convenient to work in the spatially-flat gauge, where the scalar-type perturbative degrees of freedom are encoded in the scalar field perturbations $Q^I = \delta\phi^I$. The quadratic action for the $Q^I$ can be obtained by expanding the action $S$, given in (\ref{action}), to quadratic order in the perturbations. Using the conformal time $\eta=\int dt/a(t)$ instead of the cosmic time $t$,  the action for the linear perturbations,   expressed in terms of the canonically-normalized variables $v^I=aQ^I$, is given by (see e.g. \cite{Langlois:2010xc})
    \begin{equation}
S^{(2)}=\frac{1}{2}\int d\eta \, d^{3}x\left[v'^{T}v'-\partial_{i}v^{T}\partial_{i}v-a^{2}v^{T}\left(\bm{M}-H^{2}\left(2-\epsilon\right)\right)v\right],\label{S_canon}
\end{equation}
where we use a matrix notation, $v$ being the column vector with components $v^I$ and $v^T$ is the corresponding transposed matrix. A prime denotes a derivative with respect to the conformal time $\eta$. The matrix  $\bm{M}$ corresponds to  the (squared) mass matrix and is given explicitly by
    \begin{equation}
        M_{IJ}\equiv V_{,IJ}+\left(3-\epsilon\right)\dot{\phi}_{I}\dot{\phi}_{J}+\frac{1}{H}\left(V_{,I}\dot{\phi}_{J}+\dot{\phi}_{I}V_{,J}\right).\label{M_IJ_ori}
\end{equation}
This result has also been extended to models with non-canonical kinetic terms  (see e.g. \cite{LRPST}).

If we go to  another orthonormal basis $e^I_m$, as shown in (\ref{change_basis}),  the linear perturbations will be described by  new components $u^m$, defined as
    \begin{equation}
        v^I = e^I_m u^m,\qquad \qquad m=1,\cdots,N\,.\label{modes_trans}
    \end{equation}
    The equation of motion for the $u^m$ can be derived from  the quadratic action
    \begin{equation}
        S=\frac{1}{2}\int d\eta\,  d^{3}x\left[u'^{T}u' + u^{T}\partial^2u-a^{2}u^{T}\left(\tilde{\bm{M}} -H^{2}\left(2-\epsilon\right)\right)u+2au'^{T}\bm{Z}u\right],
        \label{S_mass}
        \end{equation}
with
    \begin{equation}
        \tilde{\bm{M}} := \bm{M} + \bm{Z}^2,
    \end{equation}
where
$\bm{M}$ now denotes the matrix of coefficients  $M_{mn}$ (we use the same notation $\bm{M}$ for simplicity) defined by
    \begin{equation}
        M_{mn}:=M_{IJ}e^I_me^J_n ,
    \end{equation}
and where $\bm{Z}$ is the antisymmetric matrix of components $ Z_{mn}$ introduced in ({\ref{Z_mn_def}).
In matrix notation, these equations of motion are given by
    \begin{equation}
        u''-\partial^2u+a^2\left(\tilde{\bm{M}}-H^2(2-\epsilon) +\dot{\bm{Z}}+H\bm{Z}\right)u+2a\bm{Z}u'=0.\label{eom_full}
    \end{equation}

So far, our new basis remains arbitrary. The usual approach is to introduce a  basis that consists of unit vectors defined along the so-called    adiabatic and entropic directions, as first  introduced  in \cite{Gordon:2000hv} and further developed  in \cite{GNvT,Peterson:2011yt,LV}. The idea is to use the prefered direction in field space defined by the instantaneous background velocity $\dot{\phi}^I$ in order to project perturbations parallely and perpendicularly to the background velocity $\dot{\phi}^I$, which yields the adiabatic and entropic modes respectively. An advantage of this choice is that  the comoving curvature perturbation, which is directly relevant for observations, is  related with the adiabatic mode.
Below, we will use of a different basis, which is ideally suited to study perturbations that can be decomposed into light modes and heavy modes, as discussed in the next section.

\section{Effective theory for the light modes}{\label{sec:eff}}
In the present work, we assume that the scalar field perturbations can be divided into two categories: light modes and heavy modes. More precisely, this means that the eigenvalues
of the mass matrix $\bm{M}$, defined in (\ref{M_IJ_ori}), can be organized as follows:
    \begin{equation}
        \{ m_{1}^2,\cdots,m_{L}^2,m_{L+1}^2,\cdots, m_{N}^2 \},
    \end{equation}
with
    \begin{eqnarray}
        |m_{i}^2| \ll  H^2,\quad (i=1,\cdots L),\label{Ml}\qquad m_{j}^2  \gg  H^2,\quad (j=L+1,\cdots N)\,.\label{Mh}
        \end{eqnarray}
 The corresponding eigenvectors of the mass matrix thus define respectively the light and heavy modes.
 Note that we do not consider  here the possibility that modes with a mass of  the order of  $H$ are present among the perturbations, even if this  situation can lead to interesting effects as discussed in \cite{Chen:2009we,Chen:2009zp}. We also discard the special case where highly unstable modes (with $m^2<0$) are present.

 \subsection{Effective action for the light modes}
 It is convenient to choose an orthonormal basis that consists of the (normalized) eigenvectors of the mass matrix $\bm{M}$, which we name the {\it mass basis}.
  In this new basis, the mass matrix is by construction diagonal, i.e.
     \begin{equation}
         M_{mn}
        = \mathrm{diag}\{\cdots, m_{(l)i}^2,\cdots,m_{(h)j}^2,\cdots \}\,,
    \end{equation}
    where the subscript $(l)$ denotes the light modes and $(h)$ the heavy modes.
 The corresponding perturbation modes $u_{m}$ in (\ref{modes_trans}) can also be divided  into  light modes $u_{(l)m}$  and heavy modes $u_{(h)m}$, i.e.
    \begin{equation}
        u=\left(\begin{array}{c}
u_{(l)}\\
u_{(h)}
\end{array}\right).
    \end{equation}
    As a consequence, the antisymmetric matrix
$\bm{Z}$ can be decomposed as
    \begin{equation}
\bm{Z}=\left(\begin{array}{cc}
\bm{Z}_{(l)} & \bm{Z}_{(c)}\\
-\bm{Z}_{(c)}^{T} & \bm{Z}_{(h)}
\end{array}\right).\label{Z_split}
\end{equation}

In order to determine the effective action describing the dynamics of the light modes, we follow the strategy of  \cite{Achucarro:2010da,Achucarro:2010jv}. By neglecting the kinetic terms of the heavy modes in  (\ref{S_mass}), the variation of the action with respect to $u_{(h)}$ yields the constraints
  \begin{equation}
        -\Delta u_{(h)}+\frac{2}{a}\bm{Z}_{(c)}^{T}u_{(l)}'+\mathcal{O}^{T}u_{(l)}=0,\label{uh_eom}
        \end{equation}
with
    \begin{eqnarray}
\Delta & := & \bm{M}_{(h)}-\bm{Z}_{(c)}^{T}\bm{Z}_{(c)}-H^{2}\left(2-\epsilon\right)-\frac{\partial^{2}}{a^{2}},\label{Delta_def}\\
\mathcal{O} & := & H\bm{Z}_{(c)}+\dot{\bm{Z}}_{(c)} -\bm{Z}_{(l)}\bm{Z}_{(c)}-\bm{Z}_{(c)}\bm{Z}_{(h)}.\label{O_def}
\end{eqnarray}
Plugging the solutions of the constraints,
    \begin{equation}
        u_{(h)}=\Delta^{-1}\left(\frac{2}{a}\bm{Z}_{(c)}^{T}u_{(l)}'+\mathcal{O}^{T}u_{(l)}\right)
        \label{uh_sol}\,,
        \end{equation}
back  into the action (\ref{S_mass}) gives,  after some straightforward manipulations, the effective action for the light modes which can be expressed in the form
    \begin{equation}
        S_{\text{eff}}[u_{(l)}]=\frac{1}{2}\int d\eta d^{3}x\left[u_{(l)}'^{T}\bm{\mathcal{K}}u'_{(l)}+u_{(l)}^{T}\partial^{2}u_{(l)}-a^{2}u_{(l)}^{T}\left(\bm{\mathcal{M}} -H^2(2-\epsilon)\right) u_{(l)}+2au_{(l)}'^{T}\bm{\mathcal{Z}} u_{(l)}\right].\label{S_vl_eff}
        \end{equation}
The coefficients of the above action are explicitly given by the expressions
    \begin{eqnarray}
        \bm{\mathcal{K}} & = & 1+4\bm{\Xi}_{1},\label{K_cal_ex}\\
        \bm{\mathcal{M}} & = & \bm{M}_{(l)}+\bm{Z}_{(l)}^{2}-\bm{Z}_{(c)}\bm{Z}_{(c)}^{T}-H^{2}\left(\bm{\Xi}_{1}+\bm{\Xi}_{2}+\bm{\Xi}_{2}^{T}+\bm{\Xi}_{3}\right)+\frac{1}{a^{2}}\left[aH\left(2\bm{\Xi}_{1}+\bm{\Xi}_{2}+\bm{\Xi}_{2}^{T}\right)\right]',\label{M_cal_ex}\\
        \bm{\mathcal{Z}} & = & \bm{Z}_{(l)}+H\left(\bm{\Xi}_{2}-\bm{\Xi}_{2}^{T}\right),\label{Z_cal_ex}
        \end{eqnarray}
with
    \begin{eqnarray}
\bm{\Xi}_{1} & := & \bm{Z}_{(c)}\Delta^{-1}\bm{Z}_{(c)}^{T},\qquad\qquad\bm{\Xi}_{2}:=\frac{1}{H}\bm{Z}_{(c)}\Delta^{-1}\left(\dot{\bm{Z}}_{(c)}-\bm{Z}_{(l)}\bm{Z}_{(c)}-\bm{Z}_{(c)}\bm{Z}_{(h)}\right)^{T},\label{Xi123_1}\\
\bm{\Xi}_{3} & := & \frac{1}{H^{2}}\left(\dot{\bm{Z}}_{(c)}-\bm{Z}_{(l)}\bm{Z}_{(c)}-\bm{Z}_{(c)}\bm{Z}_{(h)}\right)\Delta^{-1}\left(\dot{\bm{Z}}_{(c)}-\bm{Z}_{(l)}\bm{Z}_{(c)}-\bm{Z}_{(c)}\bm{Z}_{(h)}\right)^{T}\,. \label{Xi123_2}
\end{eqnarray}
Note that all the $\bm{\Xi}_i$ are suppressed by $\Delta^{-1}$, which is defined in (\ref{Delta_def}). In the regime where we expect the effective theory applies, the eigenvalues of $\bm{M}_{(h)}$ should be much larger than the elements of the other matrices that appear in $\Delta$, and one can make an expansion in the small parameter $m_h^{-1}$, where $m_h^2$ corresponds to the typical value of the eigenvalues of  $\bm{M}_{(h)}$. Thus one can  write $\Delta^{-1} = \bm{M}_{(h)}^{-1} (1+\mathcal{O}(m_{h}^{-2}))$ and, at leading order, replace $\Delta^{-1}$ by  $\bm{M}_{(h)}^{-1}$   in $\bm{\Xi}_i$'s, which makes the expressions much simpler. For later convenience, we introduce the notation
    \[
        \bm{\Lambda} = \bm{Z}_{(c)}\bm{M}_{(h)}^{-1}\bm{Z}_{(c)}^{T},
        \]
 and thus write
    \[
        \bm{\Xi}_1 = \bm{\Lambda} + \mathcal{O}(m_{h}^{-4}).
    \]

The action
(\ref{S_vl_eff}) is one of our main results in this work. The effects of the heavy modes on the light modes are effectively encoded in $\bm{Z}_c$ and the dimensionless matrices $\bm{\Xi}_1$, $\bm{\Xi}_2$ and $\bm{\Xi}_3$. One notices that
 the influence of the massive modes multi-field manifests itself only when the background trajectory is turning in field space, i.e. when the mass basis undergoes a rotation, so that the coefficients $Z_{nm}$ are nonzero.

\subsection{Discussion on the sound speeds}
From (\ref{S_vl_eff}), the light modes are propagating with effective speeds given by the eigenvalues of
\beq
\bm{\mathcal{K}}^{-1} = (\bm{1}+4\bm{\Xi}_1)^{-1}=  \bm{1}-4 \bm{\Lambda} + \mathcal{O}(m_{h}^{-4})\,.
\eeq
In the following we discuss some particular examples.

\subsubsection{Single light mode}
When one assumes that there is only one light direction, the above matrices become numbers. One then finds that the light mode is characterized by a propagation speed given, up to order in $m_h^{-2}$, by
\beq
\label{cs_def}
c_s^2\approx 1-4 \sum_{j=2}^N\frac{Z_{1j}^2}{m_{(h)j}^2}+\mathcal{O}(m_h^{-4})
\eeq
where the sum is carried out over all massive modes.

In the particular case where there is only one heavy mode as well, one immediately recovers the results of \cite{Achucarro:2010da}. Indeed, since  $\bm{Z}_{(c)}$ coincides with $-\dot{\theta}$, where $\theta$ denotes the  angle of the adiabatic direction in field space which is commonly used in the studies of two-field models, one  finds that the matrix $\mathcal{K}$ reduces to a number given by
\beq
\frac{1}{c_s^2}=1+\frac{4\dot\theta^2}{m_h^2-\dot\theta^2-(2-\epsilon)H^2+\frac{k^2}{a^2}}\,,
\eeq
where $k$ is the wavenumber associated with the Fourier modes of the perturbations.

\subsubsection{Several light modes}{\label{sec:1l2h}}
When several light modes are involved, one can obtain an anisotropic propagation of the perturbations, similarly to what can be found with generalized multi-field Lagrangians with non-standard kinetic terms (see \cite{LRPST}). The corresponding sound speeds can be obtained by diagonalizing the matrix $\bm{\Lambda}$, whose explicit elements now read
    \begin{equation}
        \Lambda_{ij}=\sum_{h=L+1}^{N}\frac{Z_{ih}Z_{jh}}{m_h^2}
    \end{equation}
 where the index $h$ is summed over the heavy field indices.

 Denoting $\lambda_i$ the eigenvalues of the matrix $\Lambda$, one finds that the associated sound speeds are simply given by
    \begin{equation}
        c_{i}^2 = 1 - 4 \lambda_{i}+\mathcal{O}(m_h^{-4})\,,
    \end{equation}
which is  the direct generalization of (\ref{cs_def}).

Note that if $\sum_l u^l Z_{lh}=0$ for some non-vanishing (light) vector, then $\Lambda_{ij}$ becomes degenerate, with $u^l$ as eigenvector associated with the eigenvalue $\lambda=0$. This means that the propagation speed along the $u^l$ direction is the speed of light as usual. This is the case if a given light mode is not coupled to any of the heavy modes (i.e. if $Z_{lh}=0$ for all values of $h$).
Note also that, if there are as many light modes and heavy modes, $ \bm{Z}_{(c)}$ is a square matrix and $\det(\bm{\Lambda}) = \det(\bm{Z}_{(c)})^2/\det(\bm{M}_{(h)})$. If $\bm{Z}_{(c)}$ is degenerate, then $\lambda=0$ is an eigenvalue of $\bm{\Lambda}$.

\section{Sudden turn along the inflationary trajectory}
\label{sec:turn}
In the present section, we  focus on the possibility that  the inflationary trajectory undergoes a  turn due to the sudden change of orientation of the massive directions, and correspondingly of the light directions. It is a reasonable approximation to restrict the analysis to   a two-field description in the  two-dimensional plane spanned by the velocities  before the turn and after the turn. We will also assume that there is enough time between two sudden changes of direction that   the inflationary velocity can relax to its natural direction between two turns.

During a turn, the trajectory deviates from the bottom of the potential valley because of the centrifugal force, as illustrated in Fig.\ref{fig:turn}. If the turn is sufficiently soft then the trajectory smoothly relaxes to the minimum of the potential after the turn. However, if the turn is sharp, the trajectory strongly deviates from the bottom of the valley and then relaxes via oscillations around the ``minimal'' trajectory. In the latter case,  the heavy fields become excited and   the amplitude of the oscillations depends on the ``sharpness" (defined precisely below) of the turn.
\begin{figure}[h]
\begin{minipage}{0.95\textwidth}
    \centering
    \begin{minipage}{0.6\textwidth}
    \centering
        \includegraphics[width=\textwidth]{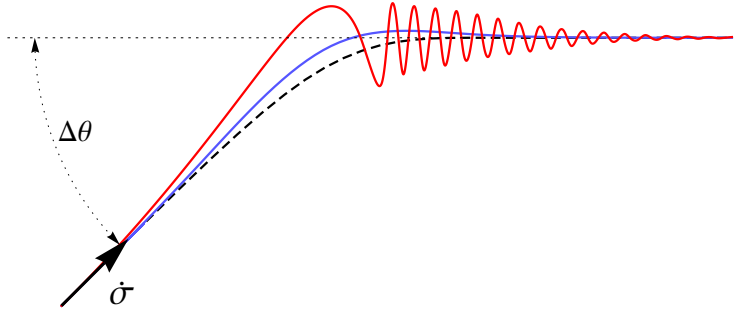}
            \caption{Schematic representation of the background trajectory in field space during a single turn. The dashed line represents the light direction of the potential (i.e. the bottom of the valley). If the turn is ``soft", the trajectory deviates slightly from the bottom and remains smooth (blue line). If the turn is sharp, the trajectory  deviates from the light valley significantly and oscillates after the turn (red line).}
            \label{fig:turn}
    \end{minipage}
\end{minipage}
\end{figure}

When the relaxation is oscillatory,  the direction of the inflationary velocity strongly varies  during the process. This  indicates that choosing the adiabatic-entropic basis to describe the inflationary evolution might not be the most appropriate as this basis undergoes wild oscillations during the turning phase. This is why  we will use the potential basis, which does not oscillate during this process, to study in detail the background trajectory during and just after the turn. The evolution of the linear perturbations will be investigated in the next section.

\subsection{Kinematic basis}
Before using the potential basis in the next subsection, let us first  describe the background evolution in terms of the usual adiabatic-entropic decomposition. The adiabatic  direction is represented by the unit vector
\beq
n^I= \frac{\dot\phi^I}{\dot\sigma}, \qquad \dot\sigma\equiv \sqrt{\delta_{IJ}\, \dot\phi^I\dot\phi^J},
\eeq
while the entropic direction is along the unit vector $s^I$, which is orthogonal to $n^I$. We will call the basis $\{n^I, s^I\}$ the ``kinematic basis'', since $n^I$ is pointing in the velocity direction.

Starting from the background equation of motion (\ref{phi_eom_pb}) in the original basis (or from (\ref{eom_bck_gen}) in an arbitrary orthonormal basis), the projection onto the adiabatic (velocity) direction, whose components are denoted $n^I$ (or  $n^m$),  yields
    \begin{equation}
    \label{bckgd_adiab}
        \ddot{\sigma} +3H\dot{\sigma} +V_{,\sigma} = 0\,, \qquad V_{,\sigma} \equiv n^I V_{,I}\,.
    \end{equation}
Since $n^I$ is a unit vector, its time variation is orthogonal to $n^I$, and therefore proportional  to $s^I$, so that one can write
\beq
\dot n^I:=\dot\theta s^I
\eeq
where the coefficient of proportionality,  $\dot\theta$, corresponds to  the time derivative of the angle $\theta$ of $n^I$ with respect to the initial field basis.

One can also derive a second order equation of motion for $\theta$, by using the projection
 of the background equation (\ref{phi_eom_pb}) along the entropic direction, which reads
 \beq
 \dot\sigma\dot\theta+V_{,s}=0, \qquad V_{,s} \equiv s^I V_{,I}\,.
 \eeq
 The time derivative of this equation yields
 \beq
 \dot\sigma\ddot\theta+\ddot\sigma\dot\theta+\dot\sigma V_{\sigma s}-\dot\theta V_{,\sigma}=0\,,\qquad V_{,\sigma s} \equiv  n^I s^J V_{,IJ}\,,
 \eeq
 where we have used $\dot s^I=-\dot\theta n^I$.
Dividing by $\dot\sigma$ and  using the adiabatic equation (\ref{bckgd_adiab}) to eliminate $V_{,\sigma}$, we finally obtain
\begin{equation}{\label{theta_2nd_eom_gen}}
        \ddot{\theta}+\left(3H+2\frac{\ddot{\sigma}}{\dot{\sigma}}\right)\dot{\theta}+V_{,\sigma s}=0\,.
    \end{equation}

\subsection{Potential basis}
We now restrict ourselves to two-field models, working in an arbitrary orthonormal basis where the components of the adiabatic and entropic unit vectors are denoted $n^m$ and $s^m$, respectively.  In this case, the entropic subspace is one-dimensional,  along  $s^m$ which can be expressed as
\begin{equation}
        s^m = -\epsilon_{mn} n^n,
    \end{equation}
    where $\epsilon_{mn}$ is the fully antisymmetric two-dimensional tensor (with $\epsilon_{12} =1$).
This implies in particular
    \begin{equation}
        V_{,\sigma s} = n^m s^n V_{,mn}=n_1 n_2 \left( V_{,22} - V_{,11} \right) + \left( n_1^2 - n_2^2 \right) V_{,12},
    \end{equation}
    where the components of the adiabatic unit vector can be written in terms of an angle $\psi$
     \begin{equation}{\label{psi_def}}
        \{n_1,n_2\} = \{\cos\psi,\sin\psi\}\,.
    \end{equation}

So far, we have been working with an arbitrary orthonormal basis. It is now convenient, when there exists a mass hierarchy between the various directions in field space as assumed in the present work, to choose  explicitly  the ``potential'' basis, i.e. the eigenvectors of the Hessian matrix of the potential.
 In the potential basis, we thus have
     \begin{equation}{\label{V_mn_diag}}
        V_{,mn} = \mathrm{diag}\{\m_{l}^2, \m_{h}^2\},\qquad\text{with } \quad \m_l\ll H\ll \m_h.
    \end{equation}
Note that the eigenvalues of the Hessian of the potential,   $\m_{l}^2$ and  $\m_{h}^2$, differ from the eigenvalues of the mass matrix, introduced earlier. In practice, the heavy eigenvalues often coincide because the extra terms in the definition (\ref{M_IJ_ori}) usually provide only small corrections. 

    The angle  $\psi$ introduced in (\ref{psi_def}) corresponds to the angle of the background velocity with respect to the ``potential basis". The angle $\theta$ is thus the sum
 \beq
 \theta = \psi +\theta_p,
 \eeq
 where $\theta_p$ is the angle between the potential basis and the original field basis.
 Substituting this decomposition into (\ref{theta_2nd_eom_gen}), one obtains
    \begin{equation}{\label{theta_2nd_eom_2f}}
        \ddot{\psi}+ 3H\left(1+\frac{2\ddot{\sigma}}{3H\dot{\sigma}}\right)\dot{\psi}+ \frac{1}{2} \left(\m_h^2-\m_l^2 \right)\sin(2\psi)= - \ddot{\theta}_p - 3H\left(1+\frac{2\ddot{\sigma}}{3H\dot{\sigma}}\right)\dot{\theta}_p\,,
    \end{equation}
    i.e. an equation of motion for $\psi$, where the source term depends on the angle $\theta_p$.
    The interest of this equation is that, in general, one expects the behaviour of $\theta_p$ to be rather smooth whereas the angles $\psi$ and $\theta$ will oscillate wildly if the inflationary trajectory is strongly displaced from the local minimum of the potential. With the above equation, one can simply interpret the evolution of $\psi$ as a response to the change in the potential landscape.

In order to obtain analytical solutions for (\ref{theta_2nd_eom_2f}), we will assume that the inflationary trajectory satisfies the slow-roll approximation, so that  $\ddot{\sigma}/(H\dot{\sigma})$ can be neglected. We also assume that the angle $\psi$ is sufficiently small so that the sinus can be replaced by its argument. With these approximations, (\ref{theta_2nd_eom_2f}) reduces to
    \begin{equation}{\label{psi_eom_sim}}
        \ddot{\psi}+ 3H\dot{\psi}+ \m_h^2\psi = - \ddot{{\theta}}_p - 3H\dot{{\theta}}_p,
    \end{equation}
where one recognizes the familiar equation of motion  of a damped harmonic oscillator with a source term that depends  on ${\theta}_p(t)$.
Assuming that $H$ and $\m_h$ remain approximately constant during the turn, (\ref{psi_eom_sim}) can be formally  solved as
    \begin{equation}{\label{psi_sol_gf}}
        \psi (t)= -\int^t dt'\, G(t,t') \left[ \ddot{{\theta}}_p(t') + 3H\dot{{\theta}}_p(t') \right],
    \end{equation}
where the retarded Green's function is given by
    \begin{equation}{\label{psi_green_fun}}
        G(t,t') = \Theta(t-t') \frac{\sin (\omega (t-t'))}{\omega} e^{-\frac{3}{2}H(t-t')}   ,\qquad\qquad \omega = \sqrt{\m_h^2- \frac{9}{4}H^2},
    \end{equation}
    where $\Theta$ is the Heaviside distribution.

\subsection{Modelling the turn}{\label{sec:turn_ana}}
If  the potential is sufficiently regular, the evolution of the angle $\theta_p(t)$ during the turn can be described by the expression
    \begin{equation}
        {\theta}_p (t)= \Delta\theta \, S(t-t_0),
    \end{equation}
where $\Delta\theta$ is the global variation of the angle during the turn,
$t_0$ represents the turning time, and $S(t)$ is some function which interpolates  smoothly from 0 to 1.

The precise shape of the function $S(t)$ depends on the details of the potential, but one can try to understand the generic features of the behaviour  of $\psi$ by introducing a simple Gaussian shape for the time derivative of $S(t)$, so that
    \begin{equation}
    \label{thetapdot}
        \dot{{\theta}}_p(t) = \Delta\theta\frac{\mu }{\sqrt{2\pi}} e^{-\frac{1}{2}\mu^2t^2}\,.
    \end{equation}
We have fixed $t_0=0$ for simplicity and the parameter   $\mu$ characterizes the duration of the turn: $\Delta t_{\rm turn}\sim \mu^{-1}$. We thus have three characteristic energy scales: $H$,
$\omega (\approx \m_h \gg H)$ and $\mu$.

Plugging (\ref{thetapdot})  into (\ref{psi_sol_gf}), we get
    \begin{eqnarray}
\psi(t)
 & = & -\frac{\Delta\theta}{2} \sqrt{1+\frac{9H^2}{4\omega^2}} e^{-\frac{3}{2}Ht}\, \Re\left[e^{i\alpha+\varphi^{2}/2}e^{-i\omega t}\mathrm{erfc}\left(-\frac{\mu t-\varphi}{\sqrt{2}}\right)\right],\label{psi_ana}
\end{eqnarray}
where $\Re$ denotes the real part of the argument and $\mathrm{erfc}(z) = 1- \mathrm{erf}(z)$ is the  complementary error function. We have also introduced the parameters
    \begin{equation}{\label{alpha_varphi_def}}
        \alpha := \arctan\left( \frac{3H}{2\omega} \right),\qquad\varphi := \frac{\omega}{\mu}\sqrt{1+\frac{9H^{2}}{4\omega^{2}}}\, e^{i\left(\frac{\pi}{2}-\alpha\right)},
    \end{equation}
Note that, since $\omega\gg H$, we have $\alpha\approx 3H/(2\omega)\ll 1$ and $\varphi\approx i \omega/\mu$.

It is now useful to distinguish two distinct qualitative behaviours depending on the value of the ratio $\mu/\omega$:

\subsubsection{Sharp turn ($\mu/\omega\gtrsim 1$)}

In this case, $\psi(t)$ can be approximated by
            \begin{equation}{\label{psi_sharp_app}}
                \psi(t) \approx -\frac{\Delta\theta}{2}e^{-\frac{\omega^{2}}{2\mu^{2}}}\mathrm{erfc}\left(-\frac{\mu t}{\sqrt{2}}\right)e^{-\frac{3}{2}Ht}\cos\left(\omega t-\alpha-\frac{3H\omega}{2\mu^{2}}\right).
            \end{equation}
The evolution of $\psi(t)$ is illustrated in Fig.\ref{fig:psi_sharp_app}.
        \begin{figure}[h]
            \centering
            \begin{minipage}{0.65\textwidth}
            \includegraphics[width=0.9\textwidth]{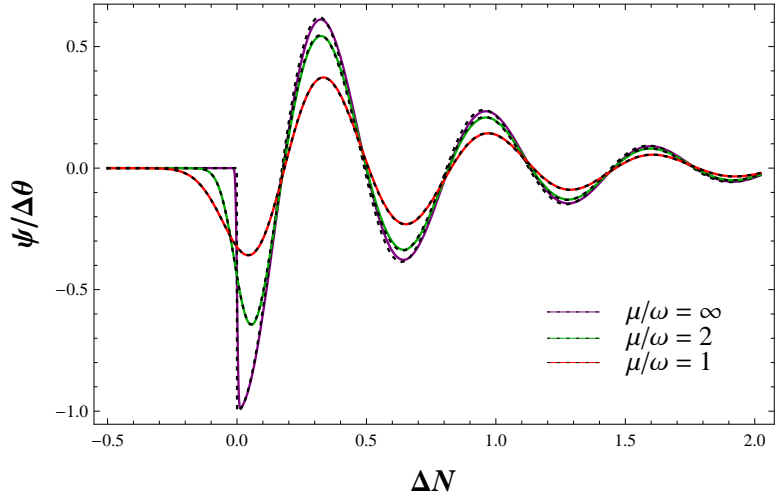}
                \caption{Evolution of the angle $\psi$ during a sharp turn as a function of the number of e-folds, for various values of $\mu/\omega$ (and $m_h/H=10$). The colored continuous lines correspond to the  numerical integration of Eq.(\ref{theta_2nd_eom_2f}) with constant $H$, while the black dashed lines are obtained from  the analytical approximation Eq.(\ref{psi_sharp_app}).}
                \label{fig:psi_sharp_app}
                \end{minipage}
            \end{figure}
Long  before the turn, $-\mu t\gg 1$, $\mathrm{erfc}(-\mu t/\sqrt{2})\rightarrow 0$ and thus $\psi\rightarrow 0$. Soon after the turn, $\mathrm{erfc}(-\mu t/\sqrt{2})\rightarrow 2$, and $\psi$ starts to oscillate with the damping factor $e^{-\frac{3}{2}H t}$. Note that  the initial amplitude of oscillation is approximately $\sim e^{-\frac{\omega^{2}}{2\mu^{2}}} \Delta \theta <\Delta \theta$.

In the limit $\mu\gg\omega$, the expression (\ref{psi_sharp_app}) further simplifies into
 \begin{equation}{\label{psi_sol_sharp_an}}
        \psi(t) = -\Delta\theta \, e^{-\frac{3H}{2}t} \cos\left(\omega t -\alpha\right)\,\Theta(t)\, ,
    \end{equation}
which can also be obtained directly from (\ref{psi_sol_gf}) by assuming  an {\it instantaneous} source term, i.e. $S(t) = \Theta(t)$ and therefore $\dot{{\theta}}_p(t) = \Delta\theta \, \delta(t)$. For an instantaneous turn, the amplitude of $\psi(t)$ is maximal  at the turn and then oscillates with an amplitude damped by   Hubble friction.

\subsubsection{Soft turn ($\mu /\omega\ll 1$)}
    In this limit, \emph{before} and \emph{during} the turn, (\ref{psi_ana}) can be approximated by
        \begin{equation}{\label{psi_soft_app}}
            \psi(t) \approx \frac{\Delta\theta}{\sqrt{2\pi}}\,\frac{\mu^2}{\omega^{2}}\, e^{-\frac{1}{2}\mu^{2}t^{2}}\left( \mu t-3\frac{H}{\mu}\right)\,,
        \end{equation}
which shows    that there is no oscillating response during a soft turn, as illustrated in Fig.\ref{fig:psi_soft_app}.
            \begin{figure}[h]
            \centering
            \begin{minipage}{0.65\textwidth}
            \includegraphics[width=0.9\textwidth]{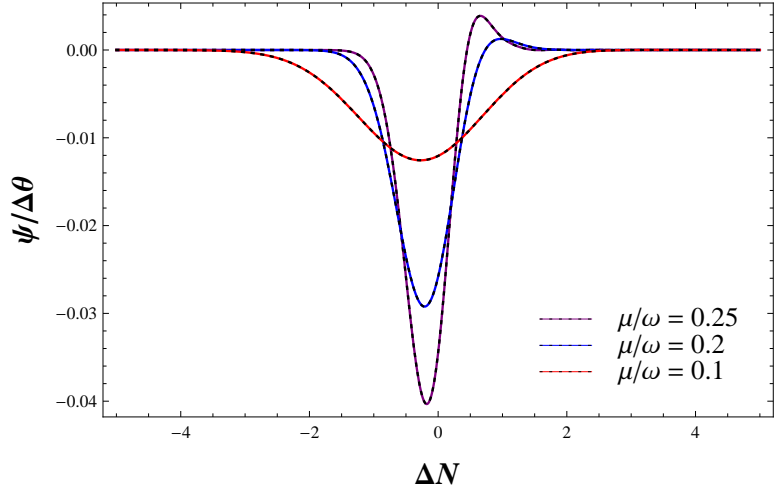}
                \caption{Evolution of $\psi$ during a soft turn (for $m_h/H=10$). The colored lines correspond to the numerical integration of Eq.(\ref{theta_2nd_eom_2f}) with constant $H$, while the  black dashed lines are given by the  analytical approximation Eq.(\ref{psi_soft_app}).}
                \label{fig:psi_soft_app}
                \end{minipage}
            \end{figure}
        The largest value of $|\psi|$ occurs around the time
            \begin{equation}
                t_{\ast}\approx -\frac{3H}{2\mu^{2}}\gamma,\qquad \gamma:=\sqrt{1+\frac{4\mu^{2}}{9H^{2}} }-1,
            \end{equation}
        which is before the turning point  $t_0=0$ and the corresponding value of $\psi$ is
            \begin{equation}
                \psi_{\mathrm{min}} = \psi(t_{\ast}) \approx -\Delta\theta\frac{3H\mu}{2\sqrt{2\pi}\omega^{2}}\left(2+\gamma\right)
                e^{-\frac{9H^{2}}{8\mu^{2}} \gamma^{2}}.
            \end{equation}
        Since $\omega\gg \mu,H$, the above amplitude is much smaller than $\Delta\theta$.

        Long \emph{after} the turn, when  $\mu t \gtrsim \omega/\mu\gg 1$, $\psi$ possesses an oscillating component but with a tiny (and thus negligible) amplitude since $\omega/\mu\gg 1$:
            \begin{equation}
                \psi\left(t\right)\approx-\Delta\theta\left[e^{-\frac{\omega^{2}}{2\mu^{2}}}e^{-\frac{3}{2}Ht}\cos\left(\omega t-\alpha-\frac{3H\omega}{2\mu^{2}}\right)-\frac{1}{\sqrt{2\pi}}e^{-\frac{1}{2}\mu^{2}t^{2}}\frac{1}{\mu t}\right].
            \end{equation}

        To summarize, in a soft turn, the background trajectory is almost the same as the light direction of the mass basis, i.e. the kinematic basis  is tightly following the mass basis. There is no explicit oscillation of the trajectory. Moreover, the ``softer" the turn is, the closer the background trajectory is to the direction of mass basis.

\subsection{Numerical study}{\label{sec:model_num}}

In order to confirm the qualitative conclusions of our analytical  approximations, we have studied numerically  a concrete two-field inflationary model with a potential that contains a valley with a turn.
Our potential for the scalar fields $\phi$ and $\chi$
is given by
\begin{eqnarray}
V(\phi,\chi) &=&
\frac12 M^2 \cos^2 \left(\frac{\Delta \theta}{2}\right)
\left[\chi - (\phi-\phi_0) \tan \Xi
\right]^2+
\frac12 m_\phi ^2 \phi^2\,,
\label{func_potential}
\end{eqnarray}
where $\Delta \theta$ is a constant parameter
corresponding to the net angle variation due to the turn.
The function $\Xi$ is defined by
\begin{eqnarray}
\Xi =  \frac{\Delta \theta}{\pi} \arctan [s (\phi - \phi_0)]\,,
\end{eqnarray}
where $s$ is a constant parameter that controls the sharpness
of the turn.
By  choosing $M \gg m_\phi$,
as illustrated in
Fig.~\ref{fig:potentialcontour_bgnm},
this potential yields a steep-sided valley,
characterized by
\begin{eqnarray}
\chi = (\phi - \phi_0)\tan \Xi\,,
\label{eq_valley}
\end{eqnarray}
which undergoes  a turn around
$\phi=\phi_0,\;\;\chi=0$.

\begin{figure}[h]
\begin{minipage}{0.45\textwidth}
    \centering
        \includegraphics[width=\textwidth]{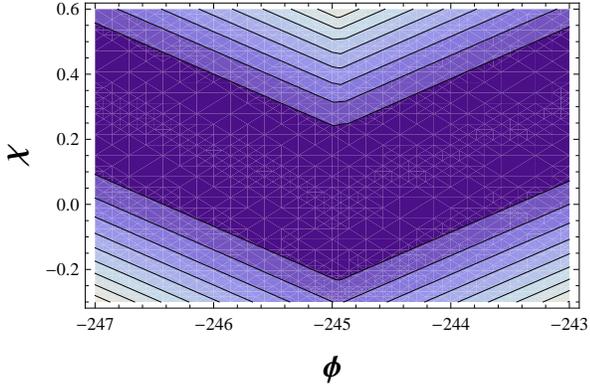}
            \caption{Contour plot of the valley corresponding to the potential Eq.~(\ref{func_potential}),
with the numerical parameters given in Eq.~(\ref{parameter_set}).
}
            \label{fig:potentialcontour_bgnm}
\end{minipage}
\end{figure}

To make easier the comparison with our analytical model of the turn, it is useful to obtain a rough  relation between the parameters $s$ and $\mu$. This can be done by approximating  the function $\arctan$ with the error function ${\rm erf}$, more precisely $\arctan(x)\approx \pi {\rm erf}(x)/2$, where the proportionality coefficient is chosen so that the asymptotic limits coincide. This yields
\begin{eqnarray}
\Xi \approx \frac{\Delta \theta}{2} {\rm erf} (s(\phi-\phi_0))
= \frac{\Delta \theta}{\sqrt{\pi}}
\int^{s(\phi-\phi_0)} _0 e^{-y^2} dy\,.
\label{xi_as_error_func}
\end{eqnarray}
Noting that when the inflationary trajectory evolves along the valley, the potential basis angle $\theta_p$ coincides with $\Xi$, one obtains the following approximation for $\dot\theta_p$:
\begin{eqnarray}
\dot{\theta}_p \simeq \dot{\Xi} \approx
\Delta \theta \frac{s \, \dot{\phi}}{\sqrt{\pi}}
e^{-s^2 (\phi-\phi_0)^2}\,.
\label{thetapdot_rough}
\end{eqnarray}
Comparing this expression with Eq.~(\ref{thetapdot}), we find the relation
\begin{eqnarray}
\mu \approx  \sqrt{2} \dot{\phi} s\,,
\label{rel_b_mu_1}
\end{eqnarray}
if $\dot{\phi}$ is expressed as a linear function
of $t$ (note  that  $\dot{\phi} > 0$)
Furthermore, if the last term in
Eq.~(\ref{func_potential}) dominates the potential
and the slow-roll condition is satisfied for $\phi$,
$\mu$ can be expressed as
\begin{eqnarray}
\mu \approx \frac{2 \sqrt{3}}{3} m_\phi  s\,.
\label{rel_b_mu_2}
\end{eqnarray}

We have integrated numerially the background equations
of motion for the scalar fields $\phi$ and $\chi$
in the potential (\ref{func_potential}) with the
following parameters
\begin{eqnarray}
M = 1.0 \times 10^{-4}\,,\;\; m_\phi = 1.0 \times 10^{-7}\,,
\;\;\phi_0=-100\sqrt{6}\,,\;\;\Delta \theta = \pi/10\,,
\label{parameter_set}
\end{eqnarray}
expressed in Planck mass units (i.e. $M_{Pl}=1$)
for various  values of the sharpness parameter $s$, in particular for
$s=100\sqrt{3}$, $s=500\sqrt{3}$ and $s=1000\sqrt{3}$, which correspond to the values
 $\mu/\omega=0.2$,
 $\mu/\omega=1$ and $\mu/\omega=2$, respectively, according to Eq.~(\ref{rel_b_mu_2}).
In all cases, we start our integration at the position
$(\phi, \chi) =(-2-100\sqrt{6},\;\;2 \tan(\pi/20))$
with vanishing velocity
(the initial velocity is in fact irrelevant since inflation
quickly reaches the slow-roll regime, long before the turn).

The background trajectories in field space
around the turning point
for three different values of $s$, are shown in
Fig.~\ref{fig:trajectory_bgnm}.
The turn becomes sharper for higher values of $s$.
For $s=100\sqrt{3}$, the trajectory is smooth, but oscillations are clearly visible for
$s=500\sqrt{3}$ and $s=1000\sqrt{3}$.
This is even more conspicuous when one considers the
inflationary velocity $\dot{\sigma}$
as shown in  Fig.~\ref{fig:sigmadot_bgnm}.
For the three inflationary trajectories, we have checked
numerically that the slow-roll parameter $\epsilon$ remains
very small (even if some oscillations are observed
for $s=500\sqrt{3}$ and $s=1000\sqrt{3}$).

\begin{figure}[h]
\begin{minipage}{0.95\textwidth}
    \centering
    \begin{minipage}{0.45\textwidth}
    \centering
        \includegraphics[width=\textwidth]{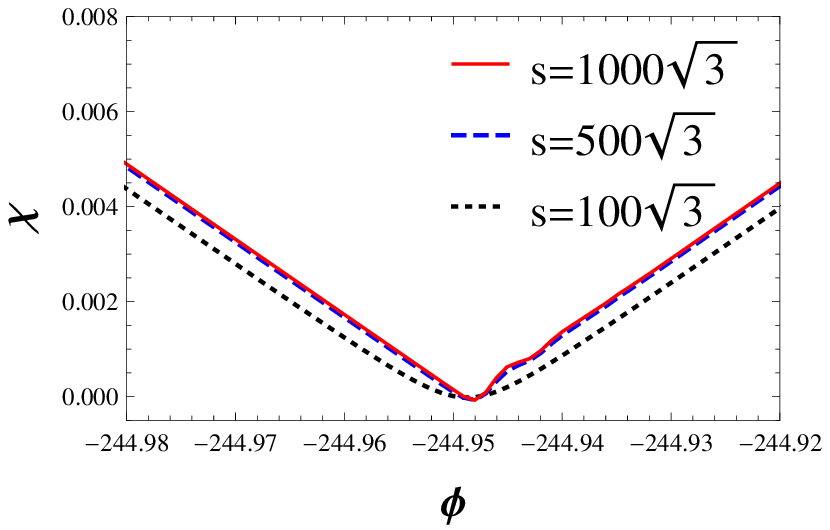}
            \caption{Background trajectories around the turn
for different values of the sharpness parameter $s$. The field $\phi$ increases with time.}
            \label{fig:trajectory_bgnm}
    \end{minipage}
    $\qquad\quad$
 \begin{minipage}{0.45\textwidth}
    \centering
        \includegraphics[width=\textwidth]{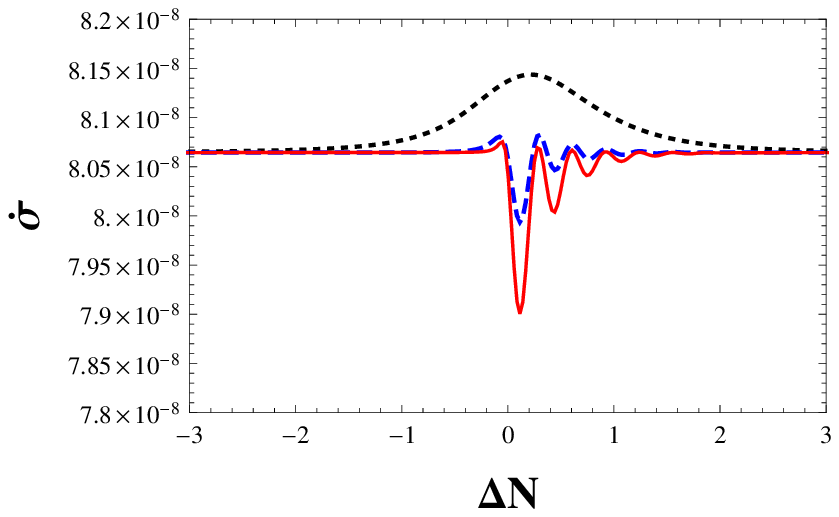}
            \caption{Evolution of $\dot{\sigma}$
around the turn for  different values of $s$.}
            \label{fig:sigmadot_bgnm}
    \end{minipage}
\end{minipage}
\end{figure}

Along the trajectory obtained numerically
one can compute the evolution of the eigenvalues
of the mass matrix which can be denoted
$m_{l}^2$ and $m_{h}^2$ as usual. One can check that the highest eigenvalue $m_h$ is of the order of $M$, as expected. For all the three trajectories considered,
the hierarchy $m_{l}^2  <  H^2 < m_{h}^2$ remains valid during the turn.
However, because of the specific shape of the potential
around the turn, the simplifying assumption that $m_h$
is constant is not strictly valid: we observe the reduction
(of about 2.5 $\%$) of $m_h$ during the turn.
Another subtlety is that $m_l ^2$ becomes negative just at the moment of the turn,
with an amplitude that increases with the sharpness
of the turn.

Let us finally discuss the evolution of the kinematic basis,
chracterized by the angle $\theta$,
with respect to that of the mass basis, which almost coincides
with the potential basis\footnote{We have checked explicitly that the potential and mass bases almost coincide: during and after the turn, we find that $(\theta_p-\theta_m)/\Delta\theta$ is at most of order $10^{-6}$.}.
For a smooth turn ($s=100\sqrt{3}$),
the velocity has the time
to adjust to the rotation of the mass basis
and the two angles evolves together
as illustrated in Fig.~\ref{fig:theta_10_bgnm}.
By contrast, for a sharp turn, the velocity follows
the light direction with some initial delay
and then oscillates, as illustrated in
Fig.~\ref{fig:theta_1000_bgnm},
for $1000\sqrt{3}$.

For comparison with our  previous discussions,
we show  the evolution of $\psi=\theta-\theta_m$
in Figs.~\ref{fig:psi_sharp_bgnm} and
\ref{fig:psi_soft_bgnm}, for the cases
with a sharp turn and a soft turn, respectively.
One sees clearly  that the evolution of $\psi$ is very
similar to that of our approximate analytical
solutions.

\begin{figure}[h]
\begin{minipage}{0.95\textwidth}
    \centering
    \begin{minipage}{0.45\textwidth}
    \centering
        \includegraphics[width=\textwidth]{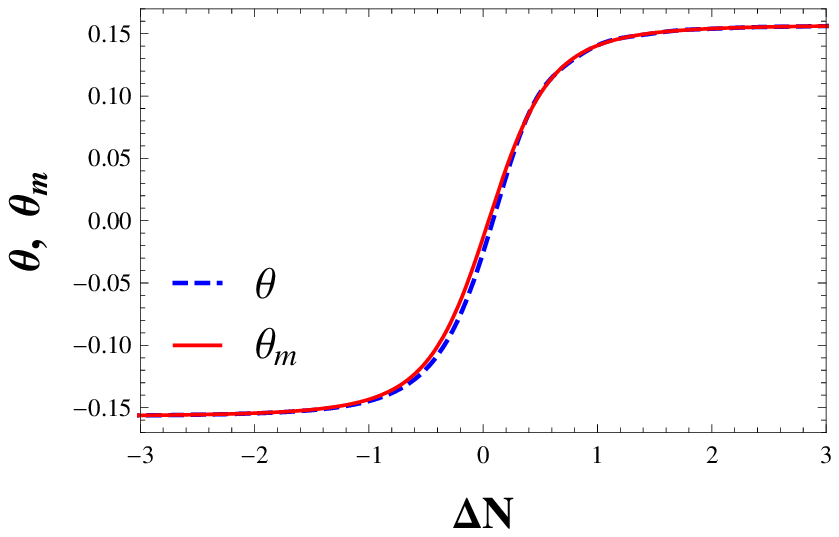}
            \caption{Evolution of $\theta$ and
$\theta_m$
around the turn for $s=100\sqrt{3}$.}
            \label{fig:theta_10_bgnm}
    \end{minipage}
    $\qquad\quad$
 \begin{minipage}{0.45\textwidth}
    \centering
        \includegraphics[width=\textwidth]{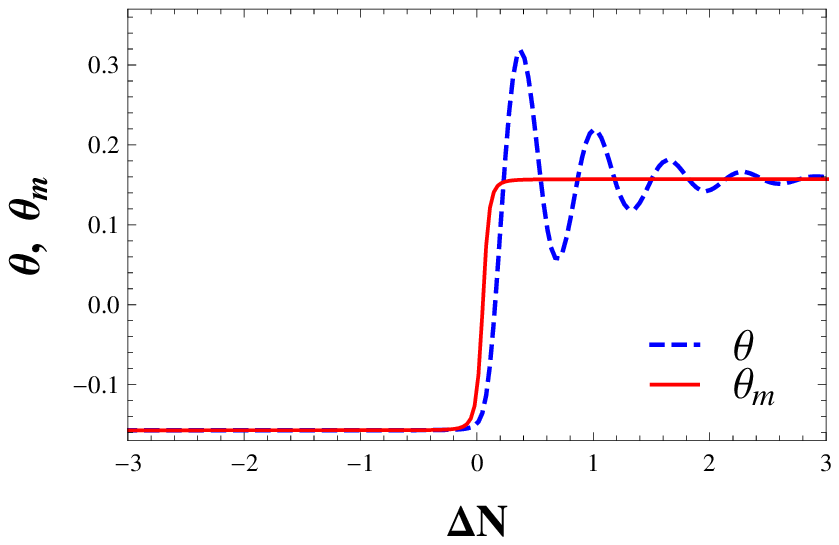}
            \caption{Evolution of $\theta$ and
$\theta_m$
around the turn for $s=1000\sqrt{3}$.}
            \label{fig:theta_1000_bgnm}
    \end{minipage}
\end{minipage}
\end{figure}

\begin{figure}[h]
\begin{minipage}{0.95\textwidth}
    \centering
    \begin{minipage}{0.45\textwidth}
    \centering
        \includegraphics[width=\textwidth]{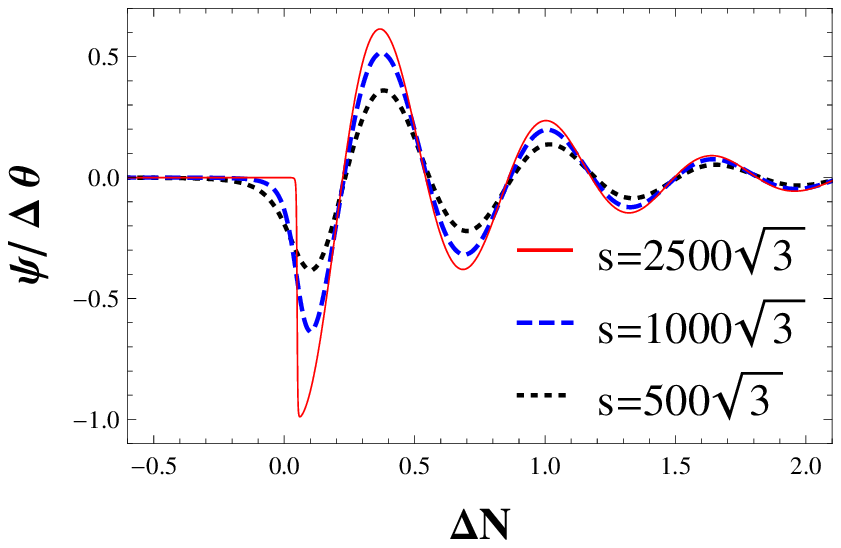}
            \caption{Evolution of $\psi$  during a sharp turn for  different values of
        $s$, corresponding to  $\mu/\omega = 1$, $2$ and $5$, respectively, according to Eq.~(\ref{rel_b_mu_2}).
}
             \label{fig:psi_sharp_bgnm}
    \end{minipage}
    $\qquad\quad$
 \begin{minipage}{0.45\textwidth}
    \centering
        \includegraphics[width=\textwidth]{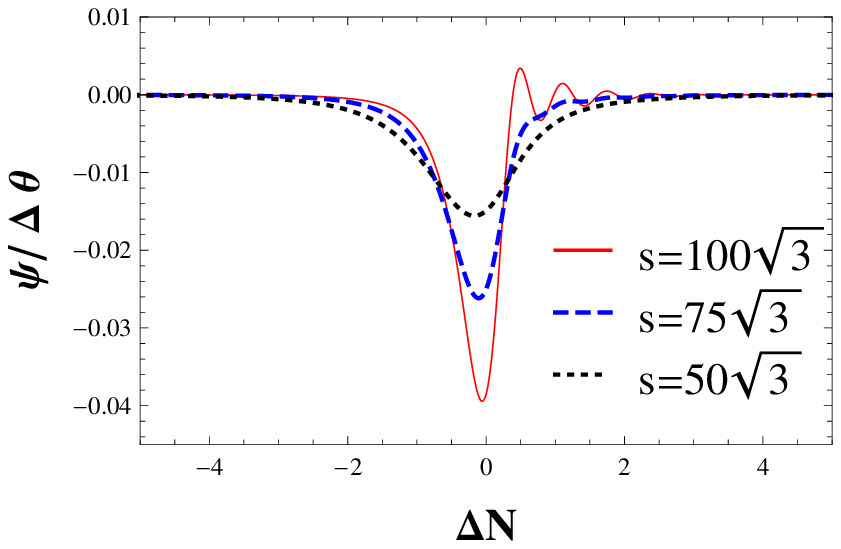}
            \caption{Evolution of $\psi$ during a   soft turn for different values of $s$, corresponding to
$\mu/\omega = 0.1$, $0.15$ and $0.2$, respectively.}
            \label{fig:psi_soft_bgnm}
    \end{minipage}
\end{minipage}
\end{figure}

\section{Evolution of the perturbations during a turn}
\label{sec:perts}

After the study of  the background evolution, we now investigate  the evolution of perturbations during the turn in order to  see whether specific signatures will be imprinted in the final power spectrum.

\subsection{Two-dimensional analysis}

\subsubsection{Adiabatic-entropic basis versus mass basis}
Let us start with the familiar approach where perturbations are described in  the adiabatic-entropic decomposition, i.e. the kinematic basis.
The equations of motion for the adiabatic and entropic modes read~\cite{Gordon:2000hv}, in Fourier space,
\begin{eqnarray}
u_{\sigma}''+k^{2}u_{\sigma}-\frac{z''}{z}u_{\sigma} & = & 2\theta'u_{s}'+2\frac{\left(z\theta'\right)'}{z}u_{s},\label{u_sigma_eom_2f}\\
u_{s}''+k^{2}u_{s}+a^{2}\left(V_{,ss} -\dot{\theta}^{2} -H^{2}\left(2-\epsilon\right)\right)u_{s} & = & -2\theta'u_{\sigma}'+2\frac{z'}{z}\theta'u_{\sigma},\label{u_s_eom_2f}
\end{eqnarray}
where $z\equiv a\sqrt{2\epsilon}$ and $\theta$ is, as previously,  the angle of the adiabatic  direction in field space. As we saw in Sec.\ref{sec:turn_ana},  during a sharp turn, the trajectory deviates from the bottom of the valley, and then rapidly oscillate around the curve that follows the bottom of the valley. As a consequence, the angle $\theta$ undergoes strong oscillations just after the turn, which makes the analysis of the above system of equations somewhat complicated.
In this respect, the situation becomes  much simpler when working in the mass basis, because the evolution of the corresponding $\theta_m$ evolves much more smoothly as we have seen earlier.

It is thus interesting to study  the equations of motion for the perturbations described in the mass basis. Using (\ref{eom_full}), one immediately finds
    \begin{eqnarray}
u_{l}''+\left(k^{2}+a^{2}m_{l}^{2}-\theta_{m}'^{2}-\frac{a''}{a}\right)u_{l} & = & \theta_{m}''u_{h}+2\theta_{m}'u_{h}',\label{eom_ul_ex}\\
u_{h}''+\left(k^{2}+a^{2}m_{h}^{2}-\theta_{m}'^{2}-\frac{a''}{a}\right)u_{h} & = & -\theta_{m}''u_{l}-2\theta_{m}'u_{l}',\label{eom_uh_ex}
\end{eqnarray}
The smoother behaviour of $\theta_m$  will enable us  to obtain approximate analytical solutions for the system  (\ref{eom_ul_ex})-(\ref{eom_uh_ex}), as we will see below\footnote{Note that $\theta_m$ (and also $\theta_p$) has an oscillatory component, because the trajectory oscillates around the bottom of the valley, but this oscillatory component is in general subdominant with respect to the smooth component.}.

Note that the descriptions in the kinematic basis and in the mass basis are simply related by a rotation of angle $\psi \equiv \theta - \theta_m$, so that, at any time, the adiabatic mode, for instance, is given by
    \begin{equation}
        u_{\sigma}=\cos\psi\,  u_{l}+\sin\psi\,  u_{h}\,,
    \end{equation}
 in terms of the light and heavy modes. Moreover,  long after the turn, the angle $\psi$ relaxes to zero if we assume that there is no subsequent turn. Therefore, the light mode and the adiabatic mode will coincide in the asymptotic limit, which allow us to compute the prediction for the adiabatic power spectrum either by using the adiabatic mode, or the light mode at late time.

\subsubsection{Three ranges of scales}
Before deriving approximate solutions for the system in the next subsection, it is useful to identify the main relevant scales.
In standard inflation, it is convenient to distinguish, at a given time, sub-Hubble and super-Hubble perturbations. In the present context, it will be useful to distinguish three intervals of wavelengths, or wave numbers, because  the heavy mode introduces the new characteristic mass scale $m_h$, in addition to the usual cosmological scale $H$.

Our system (\ref{eom_ul_ex})-(\ref{eom_uh_ex}) thus appears to have two ``horizons": one is the usual Hubble horizon with comoving radius $(aH)^{-1}$, the other is the ``heavy mass horizon" with comoving radius $(am_h)^{-1}$. Since $m_h\gg H$, the mass horizon is always deep inside the Hubble horizon. During their evolution, the modes will therefore experience two ``horizon-exit" processes, as illustrated in Fig.\ref{fig:two_scales}: first the ``mass-horizon" exit, and then  the usual Hubble exit.
\begin{figure}[h]
            \centering
            \begin{minipage}{0.6\textwidth}
            \includegraphics[width=0.7\textwidth]{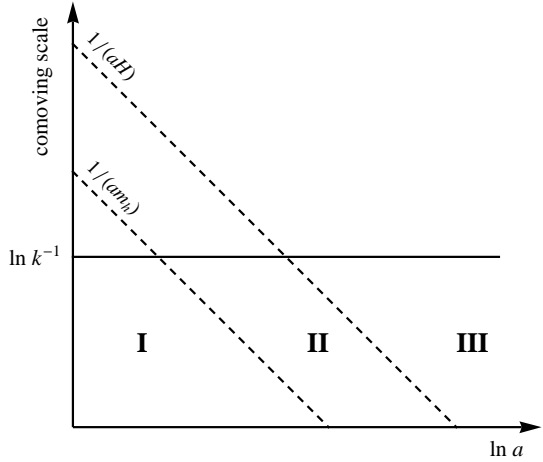}
                \caption{Evolution of the comoving  Hubble radius $(aH)^{-1}$ and of the  ``heavy mass horizon" $(am_h)^{-1}$. The Fourier modes, characterized by a fixed comoving wavenumber $k$,  exit the heavy mass horizon first and the Hubble radius later.}
                \label{fig:two_scales}
                \end{minipage}
            \end{figure}
Denoting $a_{\ast}$  the scale factor at the time of the turn, one can thus distinguish three categories of modes:
\begin{itemize}
    \item subhorizon modes $k/a_{\ast} \gg m_h$ (region I of Fig.\ref{fig:two_scales}): these modes are deep inside both the Hubble horizon and the ``mass horizon", when the turn occurs.
    \item intermediate  modes $m_h \gg k/a_{\ast} \gg H$ (region II of Fig.\ref{fig:two_scales}): these modes are inside the Hubble horizon but outside the ``mass horizon".
    \item superhorizon modes $H \gg k/a_{\ast}$ (region III of Fig.\ref{fig:two_scales}): these modes are already outside the Hubble radius at the time of the turn.
\end{itemize}
In the next subsection, we  investigate in detail  how the power spectrum is affected by the turn in the context of our simplified analytical model of Section \ref{sec:turn_ana},  considering successively the three intervals of wavenumbers discussed above.

\subsection{Approximate analytical solutions}
\label{sub_sect:approx_sols}

The system of equations  (\ref{eom_ul_ex})-(\ref{eom_uh_ex}) requires in general a numerical treatment. However, it is useful to derive approximate analytical expressions in order to gain an intuitive understanding of the qualitative behavior of the system. In order to do so, we use a perturbative approach, by noting that  the coupling between light and heavy modes is effective for a short duration only, since $\theta_m'$ is significant  only during the short time interval corresponding to the  turn.

In absence of the turn, i.e. for $\theta_m'=0$, the light and heavy modes satisfy the decoupled equations of motion
\beq
u_{l(0)}''+\left(k^{2}+a^{2}m_{l}^{2}-\frac{a''}{a}\right)u_{l(0)}=0, \qquad
u_{h(0)}''+\left(k^{2}+a^{2}m_{h}^{2}-\frac{a''}{a}\right)u_{h(0)}=0\,.
\label{system0}
\eeq
In the following, for simplicity, we will assume that $m_l$ can be neglected, $m_h$ is constant and the background evolution is approximately de Sitter. Using the usual Bunch-Davies  initial conditions, the relevant solutions for the above decoupled system are then
  \begin{eqnarray}
    u_{l(0)}\left(\eta,k\right) =  \frac{e^{-ik\eta}}{\sqrt{2k}}\left(1-\frac{i}{k\eta}\right)\label{ul0},
       \end{eqnarray}
and
    \begin{equation}
    \label{uh0}
    u_{h(0)}\left(\eta,k\right) = \frac{\sqrt{\pi}}{2}e^{-\frac{\pi}{2}\nu+i\frac{\pi}{4}}\sqrt{-\eta} \, H_{i\nu}^{(1)}\left(-k\eta\right), \qquad
        \nu\equiv \sqrt{\frac{m_{h}^{2}}{H^{2}}-\frac{9}{4}}=\frac{\omega}{H} \approx \frac{m_h}{H} \gg 1\,,
    \end{equation}
    where $H^{(1)}$ denotes a Hankel function of the first kind.

The effect of  the coupling between the light and heavy modes can be taken into account by  solving (\ref{eom_ul_ex})-(\ref{eom_uh_ex}) \emph{perturbatively} via  Green's function method.
 This means  that a  solution of the first equation in (\ref{system0}) with a source term $S$ on the right hand side can be written as   the following linear combination of positive and negative frequency modes:
    \begin{equation}
    \label{u_l}
        u_{l}(\eta,k)\simeq u_{l(0)}(\eta,k)+ C(\eta,k) \, u_{l(0)}(\eta,k)+D(\eta,k) \, u_{l(0)}^{\ast}(\eta,k),
    \end{equation}
with
    \begin{eqnarray}
C(\eta,k)  =  i\int^{\eta}d\eta'u_{l(0)}^{\ast}(\eta',k)S(\eta',k), \qquad
D(\eta,k)  =  -i\int^{\eta}d\eta'u_{l(0)}(\eta',k)S(\eta',k) \,.
\label{C_i}
\end{eqnarray}
The source term must be sufficiently small for the perturbative treatment to apply.  In the following, we will use our  simplified analytical description of the turn (\ref{thetapdot}), assuming that $\theta_p$ and $\theta_m$ coincide, and treat  $\Delta\theta$ to be a small parameter. 
On the right hand side of the  equation of motion for the light mode,  it is useful to distinguish the light contribution to the source term 
\beq
 S_l\equiv \theta_{m}'^{2}\, u_{l(0)},
 \eeq
 which is proportional to $(\Delta\theta)^2$,
 and the heavy contribution, 
    \begin{equation}
 S_h=       \theta_{m}''\, u_{h(0)}+2\theta_{m}'u_{h(0)}'\,,
    \end{equation}
    which is linear in the small parameter $\Delta\theta$. 

We are ultimately interested in the  power spectrum of the light mode on super-Hubble scales since the light mode coincides with the adiabatic mode after the relaxation phase following the turn. Because the light and heavy modes are initially statistically independent, the final power spectrum is the sum of two contributions:
\beq
{\mathcal P}(k)={\mathcal P}_{(l)}(k)+{\mathcal P}_{(h)}(k)= \lim_{k|\tau|\ll 1}\frac{k^3}{2\pi^2 a^2}\left(|u^{(l)}_{l}(\eta,k)|^2+|u^{(h)}_{l}(\eta,k)|^2\right)\,.
\eeq
 The first contribution is obtained by putting the heavy mode to zero initially while taking Bunch-Davies initial conditions for the light mode. The second contribution comes from initial conditions where the light mode is set to zero while the heavy mode is in its Bunch-Davies  vacuum. Using (\ref{u_l}), each contribution can be written as 
\beq
\label{P}
{\mathcal P}_{(l)}(k)=\left|1+C^{(l)}-D^{(l)}\right|^2 \mathcal{P}_0\,,\qquad 
{\mathcal P}_{(h)}(k)= \lim_{k|\tau|\ll 1}\frac{k^3}{2\pi^2 a^2}|u^{(h)}_{l}(\eta,k)|^2=\left|C^{(h)}-D^{(h)}\right|^2 \mathcal{P}_0
\eeq
where $C^{(l,h)}$ and $D^{(l,h)}$ denote the respective limits of the expressions (\ref{C_i}) when $\eta\rightarrow 0$, and $\mathcal{P}_0$ is the power spectrum arising from  $u_{l(0)}$, which is here scale-invariant. 

For  the spectrum due to the heavy modes, the corresponding coefficients $C^{(h)}$ and $D^{(h)}$ are obtained by plugging $S_h$ into (\ref{C_i}). This yields
\beq
{\mathcal P}_{(h)}(k)= \left| I_h\left(k\right)\right|^{2}\mathcal{P}_0 ,     \qquad  I_{h}(k) \equiv 2i \int^0 d\eta'\, \Re [u_{l(0)}\left(\eta',k\right)]\, S_{h}(\eta',k)\,,
\eeq
which is proportional  to $(\Delta\theta)^2$.
The light contribution to the spectrum contains a  zeroth order term arising from $u_{l(0)}$ and  perturbative corrections which start at quadratic order in $\Delta\theta$, since the coefficients $C_1^{(l)}$ and $D^{(l)}_1$, obtained by plugging $S_l$ into (\ref{C_i}), are already proportional to $(\Delta\theta)^2$. There is no term linear in $\Delta\theta$, but there is another term in $(\Delta\theta)^2$, which comes from the equation of motion for the heavy mode. Indeed, even if the heavy mode is initially zero, the source term on the right hand side of (\ref{eom_uh_ex}) will generate a heavy mode proportional to $\Delta\theta$. This heavy mode will in turn produce,  via the coupling on the right hand side of (\ref{eom_ul_ex}), a new contribution to the light mode which is proportional to $(\Delta\theta)^2$. This new term is more complicated to evaluate as it arises from a double integration but is  of the same order of the other contributions and must therefore be included. 

In summary, one finds that  the leading order  perturbation to the power spectrum, proportional to $(\Delta\theta)^2$ is the sum of three contributions, given by
\beq
\frac{\Delta {\mathcal P}}{{\mathcal P}_0}= {\cal F}_l+{\cal F}_h+{\cal F}_{lh}
\eeq
with
\beq
\label{F_l}
{\cal F}_l=2 \Re[I_l] \qquad        I_{l}(k) \equiv 2i \int^0 d\eta'\, \Re [u_{l(0)}\left(\eta',k\right)]\, S_{l}(\eta',k)\,,
 \eeq
\beq
\label{F_h}
 {\cal F}_h=    \left| I_h\left(k\right)\right|^{2} \qquad        I_{h}(k) \equiv 2i \int^0 d\eta'\, \Re [u_{l(0)}\left(\eta',k\right)]\, S_{h}(\eta',k)\,,
 \eeq
 and 
    \begin{eqnarray}
 {\cal F}_{lh}= 2 \Re[J_{lh}], \qquad    J_{lh}(k)&\equiv &
 2\int^{0}d\eta_{1}\Re u_{l}^{(0)}\left(\eta_{1}\right)\left[\theta_{m}''\left(\eta_{1}\right)u_{h}^{(0)}\left(\eta_{1}\right)+2\theta_{m}'\left(\eta_{1}\right)u_{h}'^{(0)}\left(\eta_{1}\right)\right]\nonumber \\
 &  & \times\int^{\eta_{1}}d\eta_{2}u_{h}^{(0)\ast}\left(\eta_{2}\right)\left[\theta_{m}''\left(\eta_{2}\right)u_{l}^{(0)}\left(\eta_{2}\right)+2\theta_{m}'\left(\eta_{2}\right)u_{l}'^{(0)}\left(\eta_{2}\right)\right]\nonumber \\
 &  & -2\int^{0}d\eta_{1}\Re u_{l}^{(0)}\left(\eta_{1}\right)\left[\theta_{m}''\left(\eta_{1}\right)u_{h}^{(0)\ast}\left(\eta_{1}\right)+2\theta_{m}'\left(\eta_{1}\right)u_{h}'^{\ast(0)}\left(\eta_{1}\right)\right]\nonumber \\
 &  & \times\int^{\eta_{1}}d\eta_{2}u_{h}^{(0)}\left(\eta_{2}\right)\left[\theta_{m}''\left(\eta_{2}\right)u_{l}^{(0)}\left(\eta_{2}\right)+2\theta_{m}'\left(\eta_{2}\right)u_{l}'^{(0)}\left(\eta_{2}\right)\right].\label{int}
\end{eqnarray}

In the following, we give some analytical expressions for the integrals $I_l$ and $I_h$, expressed in terms of $\Delta\theta$, of the  dimensionless parameters
 \beq
 \k\equiv\frac{k}{a_* H}, \qquad \tmu\equiv\frac{\mu}{H}\,,
 \eeq
as well as  $\nu$,  already defined in (\ref{uh0}). These approximate  expressions for the integrals $I_l$ and $I_h$ are obtained  by writing  $\theta'$, which appears  in the source terms $S_l$ and $S_h$, in the form
\beq
\label{thetam_prime}
\theta_m'=a\, \dot\theta_m=a_{\ast}\Delta\theta\frac{\mu}{\sqrt{2\pi}}e^{-\frac{1}{2}\mu^{2}(t-t_*)^{2}+H(t-t_*)}\,.
\eeq
The integrals $I_l$ and $I_h$ also contain the uncoupled mode functions $u_{l(0)}$ and $u_{h(0)}$, given in (\ref{ul0}) and (\ref{uh0}) respectively. In order to perform analytically the integrals $I_l$ and $I_h$, it is convenient to consider separately the three ranges of scales, where simpler expressions for $u_{l(0)}$ and $u_{h(0)}$ can be used.

\subsubsection{Scales $k/(a_{\ast}m_h)\gg 1$  (or  $x_*\gg \nu$)}

In this case, modes are  inside the ``massive horizon" (and thus also inside the Hubble horizon), which corresponds to region I in figure \ref{fig:two_scales}. Both light and heavy unperturbed mode functions are approximated  by
        \begin{equation}
            u_{l(0)} \approx u_{h(0)}\approx \frac{e^{-ik\eta}}{\sqrt{2k}}.
        \end{equation}
Assuming $x_*\gg \tmu^2 \gg 1$, in order  to get simple expressions (see the appendix for details), one finds that the dominant contributions to the light and heavy integrals are given by
\beq{\label{IlIh_case1}}
     I_l(k)\approx i\left(\Delta\theta\right)^{2}\frac{\tilde{\mu}}{4\sqrt{\pi}x_{*}}\left(e^{\frac{1}{4\tilde{\mu}^{2}}}+e^{-i\frac{\pi}{4}}\frac{\tilde{\mu}}{\sqrt{x_{\ast}}}e^{1-\tilde{\mu}^{2}+ix_{*}}\right)\,,\qquad
        I_{h}(k)\approx \Delta\theta\,  e^{\frac{1}{2\tilde{\mu}^{2}}}\,,
\eeq
where one notices that $I_h$ does not depend  on $\nu$. Since the real part of $I_l$ is suppressed by the exponential factor $\exp(1-\tmu^2)$, the contribution $\F_l$ becomes negligible while
\beq
\F_h\approx (\Delta\theta)^2\,.
\eeq

\subsubsection{Scales $H \ll k/a_{\ast}\ll m_h$ (or $1\lesssim x_* \lesssim \nu$)}
In this case, modes are outside the ``mass horizon" but still inside the Hubble radius.  In this regime, the heavy mode function (\ref{uh0}) can be replaced by
\begin{equation}
u_{h(0)}\approx \frac{e^{-i\nu(\ln(2\nu)-1)}}{\sqrt{2\nu}}e^{i\nu\ln\left(-k\eta\right)}\sqrt{-\eta}\,,\label{uh_app}
\end{equation}
where we have used the approximation~\cite{AandS}
 \beq
 H_{i\nu}^{(1)}\left(x\right)\approx\frac{2}{\Gamma(i\nu+1)}\left(\frac{x}{2}\right)^{i\nu}, \qquad \nu\gg x>0\,.
 \eeq
 as well as
 \beq
 \Gamma\left(i\nu+1\right)\approx e^{-\frac{1}{2}\pi\nu+\frac{1}{2}\ln\left(2\pi\nu\right)}e^{i\left(\frac{\pi}{4}+\nu\left(\ln\nu-1\right)\right)}, \qquad \nu\gg 1
 \eeq
for large positive $\nu$.
Using the above expression, we obtained  explicit analytical expressions for $I_l$ and $I_h(x_*, \tilde\mu, \nu)$, which are given in the appendix. As they are  rather complicated, we  discuss here, in the main text,  only the regime
 $x_{\ast}\ll \tmu$, in which the dominant contributions to the integrals reduce to  the simple expressions:
    \begin{eqnarray}
I_{l} & \approx & \frac{i\left(\Delta\theta\right)^{2}\tilde{\mu}}{2\sqrt{\pi}x_{*}^{3}}e^{ix_{\ast}}\left(x_{\ast}+i\right)\left(x_{\ast}\cos x_{\ast}-\sin x_{\ast}\right),\label{I_l_app_case2}\\
I_{h} & \approx & \frac{\Delta\theta\sqrt{\nu}}{x_{*}^{3/2}}e^{\frac{-\nu^{2}}{2\tilde{\mu}^{2}}}e^{i\nu\ln\frac{x_{\ast}}{2\nu}}\left(x_{*}\cos x_{\ast}-\sin x_{*}\right).\label{I_h_app_case2}
\end{eqnarray}
One thus finds that $I_l$ is an oscillatory function of $x_*$ (note that in the general case  discussed in the appendix, $I_l$ also contains a non-oscillatory contribution). 
The integral  $I_h$  contains an additional oscillation periodic  in  $\nu \ln x_*$ but this will disappear in the power spectrum which depends only on the modulus of $I_h$.
Concerning the amplitude of the light and heavy contributions, one can see that for large values of $x_*$ (if $\nu$ is sufficiently large), they behave as
\beq
I_l \sim (\Delta\theta)^2\frac{\tmu}{x_*}, \qquad I_h\sim\Delta\theta \, e^{-\frac{\nu^2}{2\tmu^2}}\sqrt\frac{\nu}{x_*}.
\eeq
 This implies that both amplitudes decrease for higher values of $k$, the suppression being more pronounced for $I_l$. Let us also notice that the heavy contribution is strongly suppressed when $\tmu\ll \nu$, which corresponds to a soft turn.

Substituting (\ref{I_l_app_case2}) and (\ref{I_h_app_case2}) into (\ref{F_l}-\ref{F_h}) , one finds the following  perturbations to the power spectrum:
\begin{eqnarray}
\F_l&\approx &
 - \left(\Delta\theta\right)^{2}\frac{\tilde{\mu}}{\sqrt{\pi}x_{*}^{3}}\left(x_{*}\sin x_{*}+\cos x_{*}\right)
\left(x_{*}\cos x_{\ast}-\sin x_{*}\right), \qquad 
\F_h\approx \left(\Delta\theta\right)^2\,  \frac{\nu }{x_{*}^{3}}e^{-{\nu^{2}}/{\tilde{\mu}^{2}}}
  \left(x_{*}\cos x_{\ast}-\sin x_{*}\right)^2\,.
  \label{Dk_case2_app}
\end{eqnarray}
The deformed power spectrum   is illustrated in Fig.\ref{fig:Dk} with three different choices of $\tilde{\mu}$, corresponding to increasing sharpness. The parameter $\Delta\theta=\pi/30$ is sufficiently small to find an excellent agreement between the perturbative approach and the full numerical result. As shown on the right panels of Fig.\ref{fig:Dk}, the deformation of the power spectrum is dominated by the light contribution $\F_l$ for a {\it soft turn}, but  the two other contributions $\F_h$ and $\F_{lh}$ become significant for {\it sharp turns}. Moreover, whereas $\F_h$ is necessarily positive, the terms $\F_l$ and $\F_{lh}$ can be positive or negative. 
\begin{figure}[h]
\begin{minipage}{0.98\textwidth}
    \centering
       \begin{minipage}{0.46\textwidth}
    \centering
        \includegraphics[width=\textwidth]{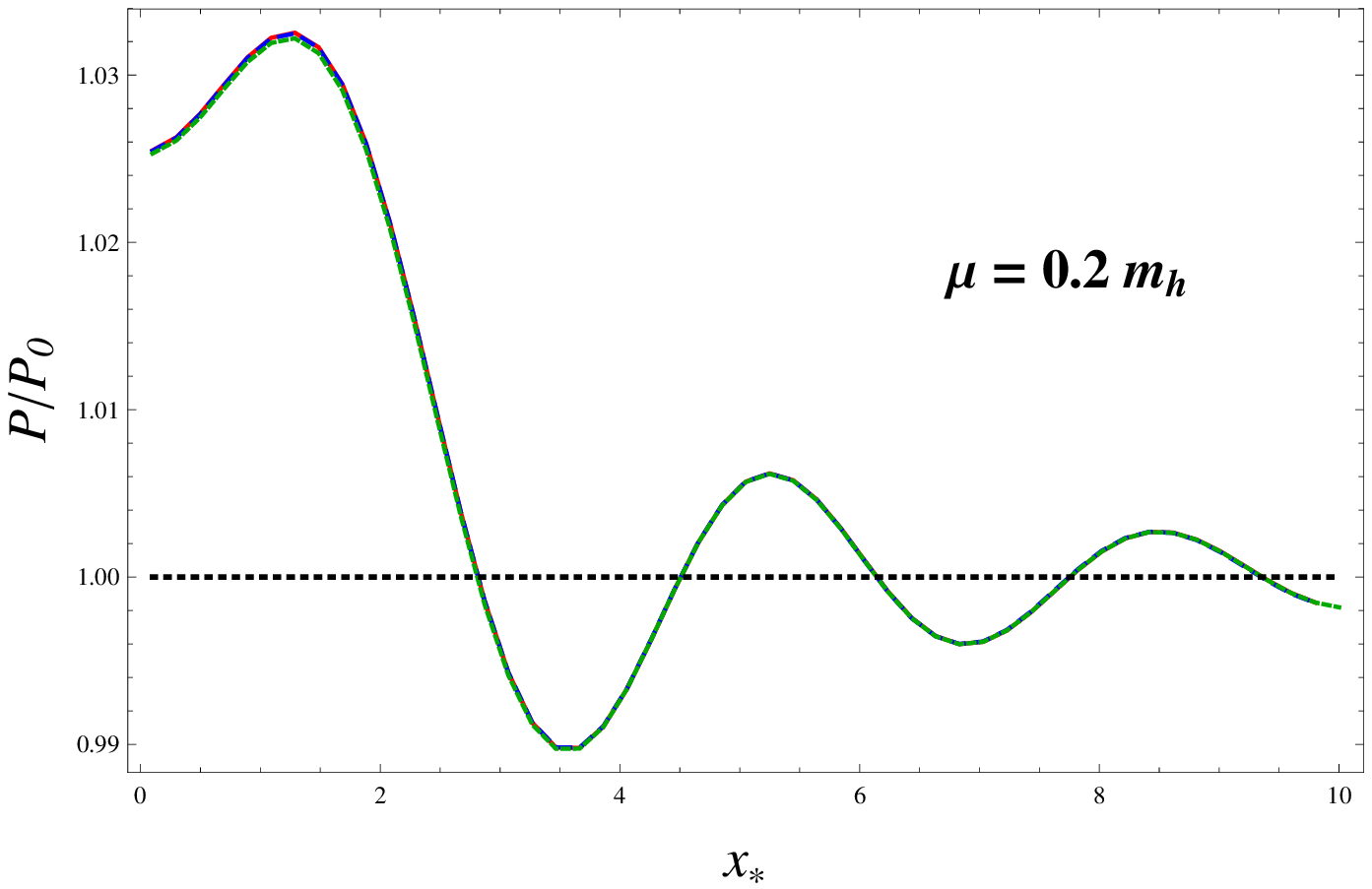}
        \end{minipage}
    $\qquad$
    \begin{minipage}{0.46\textwidth}
    \centering
        \includegraphics[width=\textwidth]{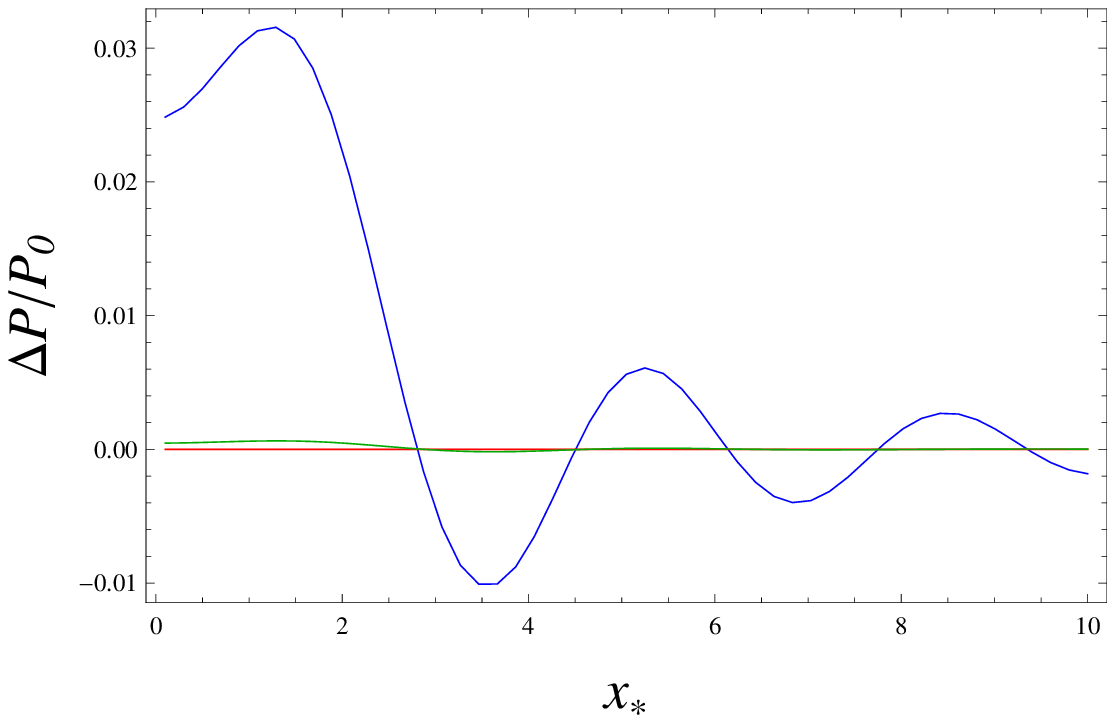}
    \end{minipage}
       \begin{minipage}{0.46\textwidth}
    \centering
        \includegraphics[width=\textwidth]{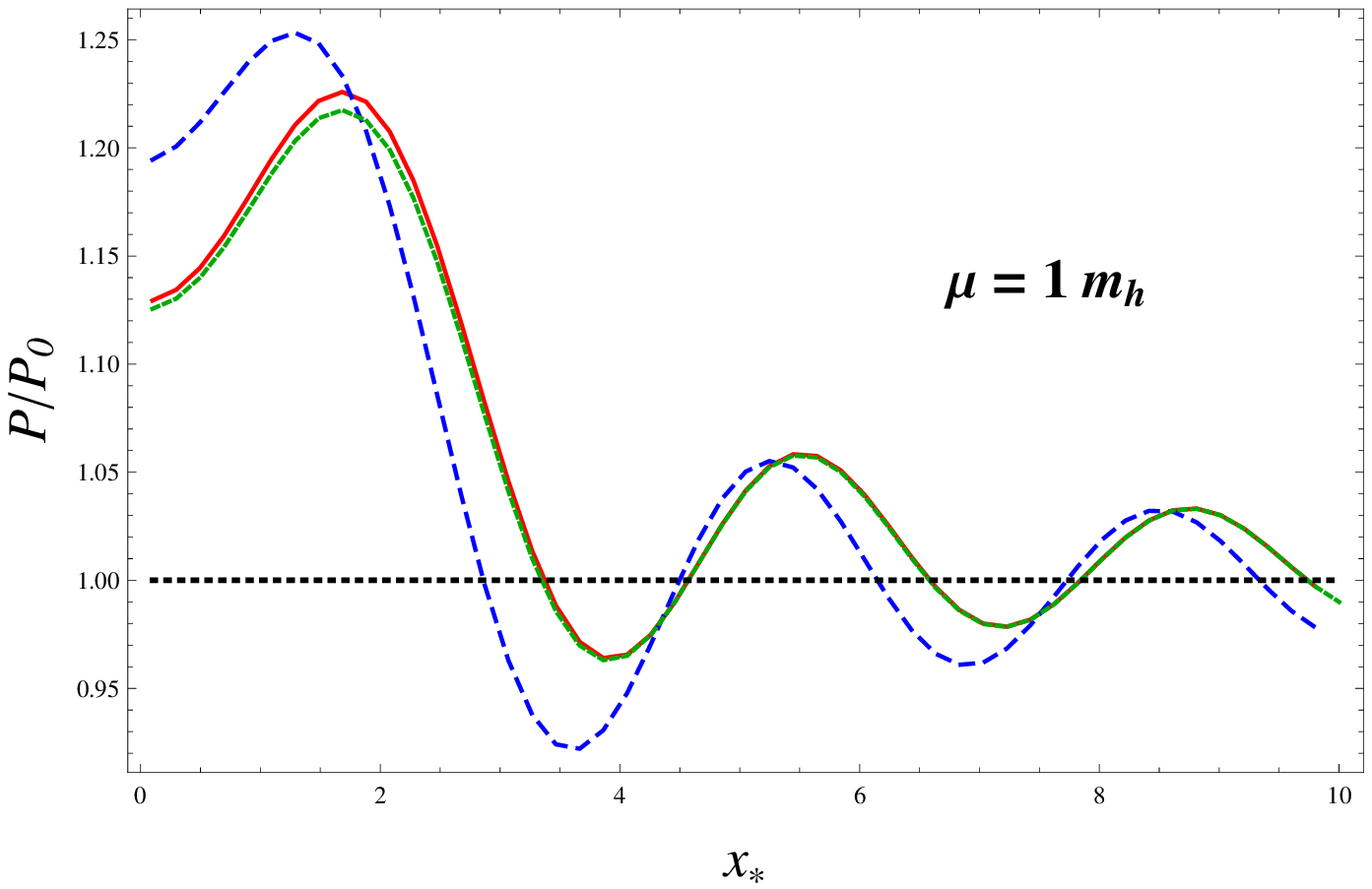}
                \end{minipage}
    $\qquad$
    \begin{minipage}{0.46\textwidth}
    \centering
\includegraphics[width=\textwidth]{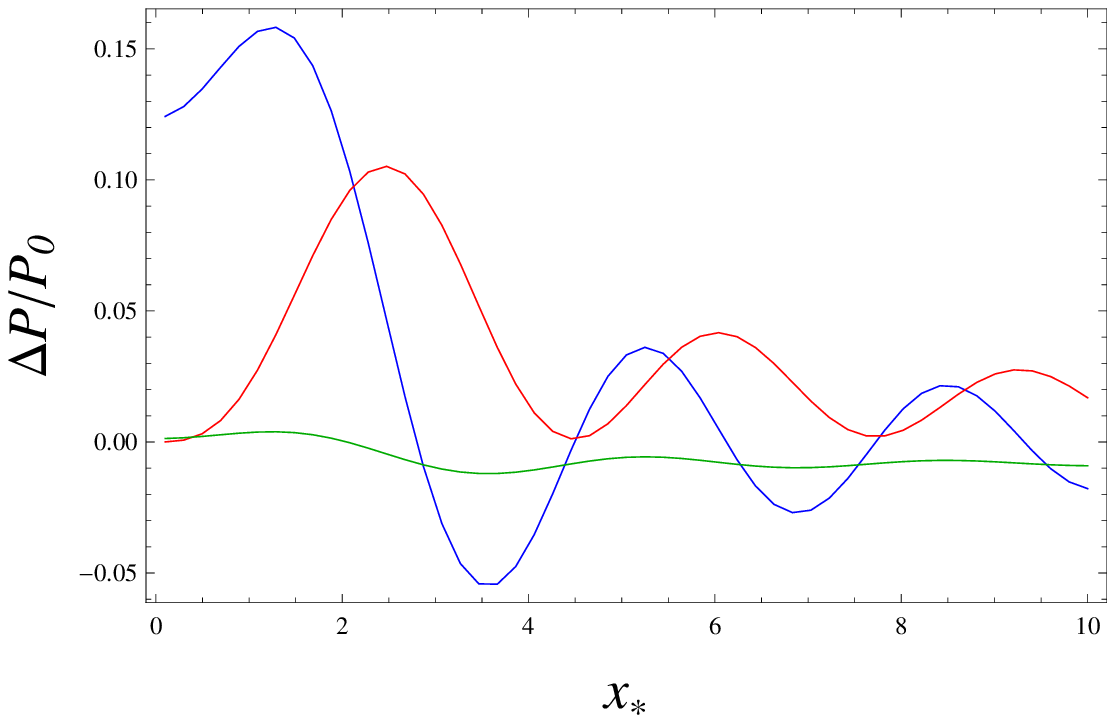}
    \end{minipage}
      \begin{minipage}{0.46\textwidth}
    \centering
        \includegraphics[width=\textwidth]{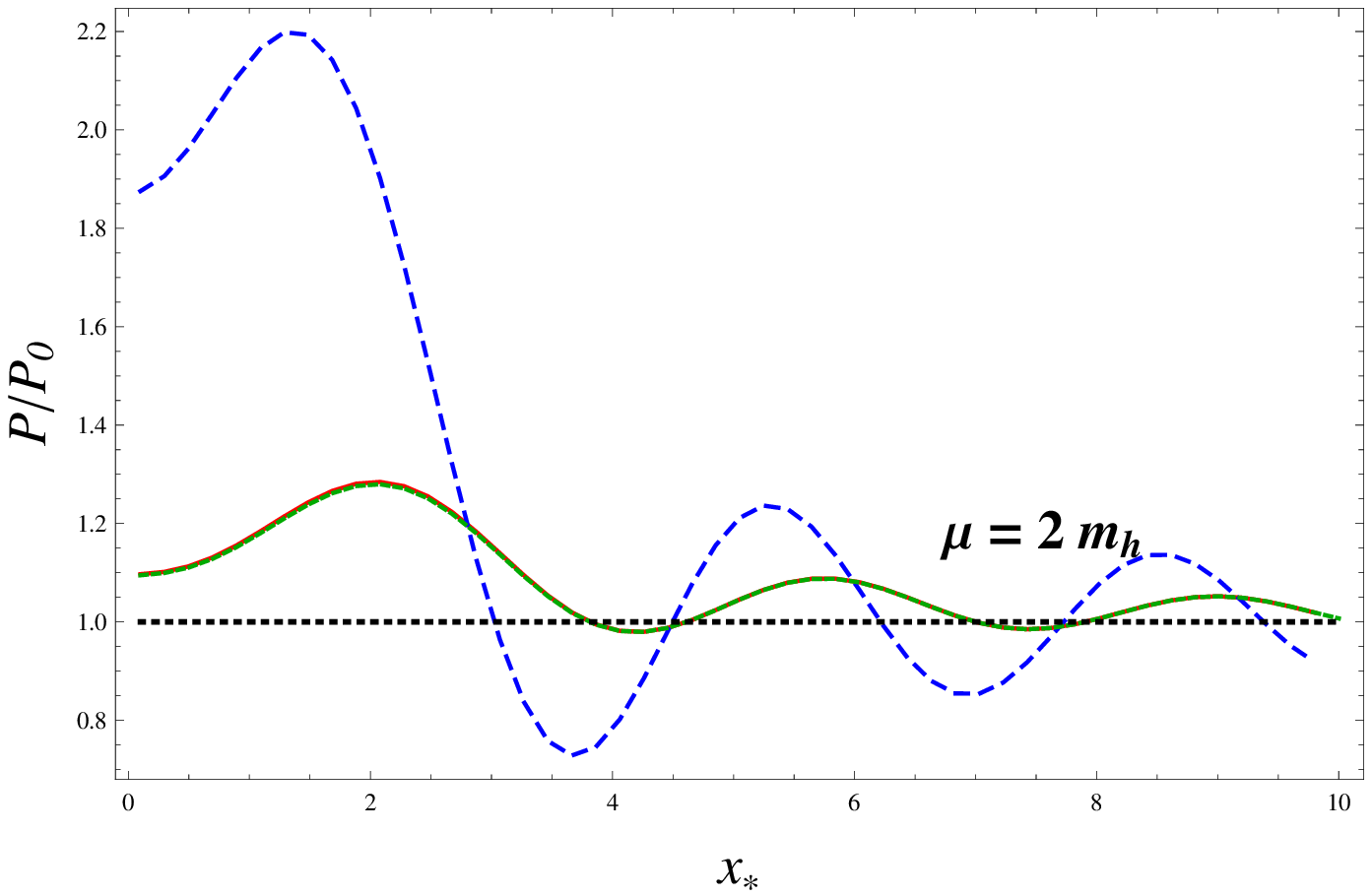}
                \end{minipage}
    $\qquad$
    \begin{minipage}{0.46\textwidth}
    \centering
\includegraphics[width=\textwidth]{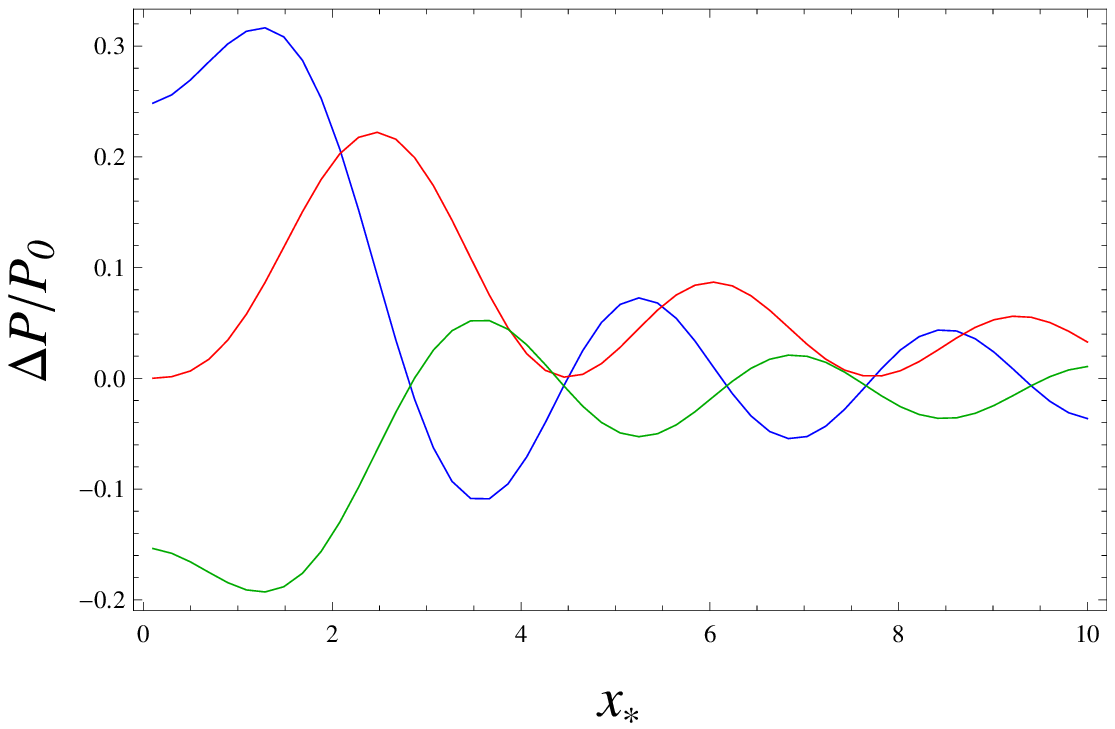}
    \end{minipage}
            \caption{Power spectra for the values  $\tmu= 12, 60, 120$, with the other fixed parameters fixed to the values $\nu=60$ and $\Delta\theta=\pi/30$. On the left panels are plotted the fully numerical result (red), the result from the perturbative approach (green) and the result from the effective theory approached (dashed blue). On the rights panels are plotted the three contributions $\F_l$ (blue), $\F_h$ (red)  and $\F_{lh}$ (green) to the perturbation of the power spectrum, i.e. the sum of these three curves (plus one) coincides with the green curves of the left panels.}
            \label{fig:Dk}
\end{minipage}
\end{figure}

\subsubsection{Scales $k/(a_{\ast}H)\ll 1$ (or $x_*\ll 1$)}
For these modes,  the turn occurs when they are already far outside the Hubble radius. One can thus obtain analytical expressions by resorting to an expansion with respect to the small parameter $x_*$, as discussed in the appendix. This leads to the integrals
  \begin{eqnarray}
I_{l} & = & \left(\Delta\theta\right)^{2}\frac{\tilde{\mu}}{18\sqrt{\pi}}\left(3+ie^{\frac{9H^{2}}{4\mu^{2}}}x_{*}^{3}\right),\label{Il_case3}\\
I_{h} & = & -\Delta\theta\frac{(2\nu+3i)}{6\sqrt{\nu}}e^{\frac{9-4\nu^2}{8\tilde{\mu}^{2}}+i\nu\left(1+\frac{3}{2\tilde{\mu}^{2}}\right)}x_{*}^{{3/2}}e^{i\nu\ln\left(\frac{x_{\ast}}{2\nu}\right)}\,.\label{Ih_case3}
\end{eqnarray}
The purely light and heavy contributions to the power spectrum perturbations are  thus given  by
        \begin{equation}{\label{Dk_case3}}
        \F_l=\frac{\left(\Delta\theta\right)^{2}}{3\sqrt{\pi}}\tmu\,,\qquad \F_h=\frac{\left(\Delta\theta\right)^{2}}{36\nu}\left(4\nu^2+9\right)
            e^{\frac{9-4\nu^2}{4\tilde{\mu}^{2}}}\, x_{\ast}^{3}\,.
        \end{equation}
The first contribution is $k$-independent term, whereas the second contribution is  proportional to $k^{3}$, which is  suppressed in the super-Hubble domain considered here.  Note that the latter term, due to the heavy modes, is  further suppressed for soft turns   via the exponential factor $\exp(-m_{h}^{2}/\mu^{2})$.

\subsection{Comparison with the effective theory}{\label{sec:pert_eff_2f}}
It is instructive to compare the results that we have just obtained in the full two-dimensional analysis with
the predictions of the effective field theory presented in Section \ref{sec:eff}. In the present case, the effective theory contains a single  mode and the equation of motion  for the corresponding canonical variable $v\equiv u_l/c_s$  is given by
    \begin{equation}{\label{eom_eff_2f}}
        v''+\left(c_{s}^{2}k^{2}+a^{2}m_{\mathrm{eff}}^{2}-\frac{a''}{a}\right)v=0,
    \end{equation}
with the effective speed of sound
    \begin{equation}
    \label{c_s}
        c_s^2 = \left( 1+4\frac{\dot{\theta}_{m}^{2}}{\hat m_h^2} \right)^{-1}\,,\qquad
          \hat m_h^2 \equiv  m_{h}^{2}-\dot{\theta}_{m}^{2}-H^{2}\left(2-\epsilon\right)+\frac{k^{2}}{a^{2}}\,,
    \end{equation}
and the effective mass
    \begin{equation}
    \label{m_eff}
        m_{\mathrm{eff}}^{2} = c_{s}^{2}\left[m_{l}^{2}-\frac{\theta_{m}'^{2}}{a^{2}}-\frac{\theta_{m}''^{2}}{\hat m_h^2 a^{4}}+\frac{2}{a^{2}}\left(\frac{1}{\hat m_h^2 a^{2}}\theta_{m}'\theta_{m}''\right)'\right]+\left(1-c_{s}^{2}\right)\frac{a''}{a^{3}}-\frac{1}{a^{2}}\left(2\frac{c_{s}'^{2}}{c_{s}^{2}}-\frac{c_{s}''}{c_{s}}\right).
    \end{equation}
In the above equations, $\theta_m$ denotes, as usual,  the angle of the mass basis with respect to  the original basis.

As before, we assume for simplicity that $H$ and $m_h$ are constant, and $m_l=0$.
In the regime where the effective theory is expected to be valid, i.e. $m_h^2\gg \dot{\theta}_m^2$, we may expand both $c_s$ and $m_{\mathrm{eff}}$ with respect to large $m_h$ to obtain
    \begin{eqnarray}
        c_s^2 & \approx& 1-\frac{4\theta_{m}'^{2}}{a^2m_{h}^{2}}+\mathcal{O}\left(m_h^{-4}\right),{\label{cs_eff_exp}}\\
        m_{\mathrm{eff}}^{2} &\approx & -\frac{\theta_{m}'^{2}}{a^{2}}+\frac{1}{a^{4}m_{h}^{2}}\left(4a^{2}H^{2}\theta_{m}'^{2}+4\theta_{m}'^{4}+12aH\theta_{m}'\theta_{m}''-3\theta_{m}''{}^{2}-2\theta_{m}'\theta_{m}'''\right)+\mathcal{O}\left(m_{h}^{-4}\right).{\label{meff_eff_exp}}
    \end{eqnarray}
In the decoupling limit where $m_h\rightarrow \infty$, one finds $c_s^2\rightarrow 1$ and $ m_{\mathrm{eff}}^{2} \rightarrow -\dot{\theta}_m^2$, and the equation of motion (\ref{eom_eff_2f}) is then equivalent to the equation of motion  for the light mode (\ref{eom_ul_ex}) with the heavy mode $u_h$  put to zero.

In order to obtain analytical solutions, we can solve (\ref{eom_eff_2f}) perturbatively via Green's function method, as we did before. For this purpose, one rewrites (\ref{eom_eff_2f}) in the form
 \beq
 {\label{eom_eff_2f_src}}
 v''+\left(k^{2}-\frac{a''}{a}\right)v=\left[\left(1-c_{s}^{2}\right)k^{2}-a^{2}m_{\mathrm{eff}}^{2}\right] \, v,
\eeq
where the right-hand side is treated as a source term. The unperturbed solution $v_{(0)}$, i.e.  the solution of the homogeneous equation (with the right-hand side put to zero), is the same as   in (\ref{ul0}). Following the same  procedure as that described at the beginning of \ref{sub_sect:approx_sols}, one finds that the perturbation of the power spectrum of $v$ is  given  by the expression
\beq
\frac{\Delta {\mathcal P}}{{\mathcal P}_0}\equiv  {\cal F}_{\rm eff} = 2 \Re[I\left(k\right)]
\eeq
with 
\beq
\label{I_eff_2f}
 I(k)\equiv 2i \int^0 d\eta'\, \Re [v_{(0)}\left(\eta',k\right)] \left[\left(1-c_{s}^{2}\right)k^{2}-a^{2}m_{\mathrm{eff}}^{2}\right]v_{(0)}\,,
 \eeq
 where $c_s^2$ and $m_{\rm eff}^2$ are explicitly given in (\ref{c_s}) and (\ref{m_eff}), respectively.

Using the Gaussian ansatz (\ref{thetam_prime}) for the turn,
 the deformation of the power spectrum is found to be (see Appendix  for details)
     \begin{equation}
\F_{\mathrm{eff}}\left(k\right) \approx \F_{0}\left(k\right)+\F_{1}\left(k\right)+\mathcal{O}\left(m_{h}^{-4}\right),\label{D_eff}
\end{equation}
where $\F_0$ is the contribution corresponding to  the decoupling ($m_h\rightarrow \infty$) limit.
For large values of $\tmu$, namely $\tmu\gg \x$, it is given by
\begin{eqnarray}
\F_{0}(k)  = \frac{\left(\Delta\theta\right)^{2}\tmu}{\sqrt{\pi} \, x_{*}^{3}} \left(x_{*}\sin x_{*}+\cos x_{*}\right)\left(\sin x_{*}-x_{*}\cos x_{*}\right)\,.
 \label{Dk_eff_0}
\end{eqnarray}
 The next order term $\F_1$, which is the main contribution due to the heavy mode, is given (again in the limit $\tmu\gg \x$)
 by  
    \begin{eqnarray}
\F_{1}(k)  =  \frac{\left(\Delta\theta\right)^{2}\tmu^3}{2\sqrt{\pi}\, x_{*}^{3}}\frac{H^{2}}{m_{h}^{2}}
  \left(x_{*}\sin x_{*}+\cos x_{*}\right)\left(\sin x_{*}-x_{*}\cos x_{*}\right)\label{Dk_eff_1}\,.
\end{eqnarray}

The behavior of the power spectrum is illustrated in  Fig.\ref{fig:Dkeff}  for different choices of parameters.
\begin{figure}[h]
\begin{minipage}{0.9\textwidth}
    \centering
    \begin{minipage}{0.46\textwidth}
    \centering
        \includegraphics[width=\textwidth]{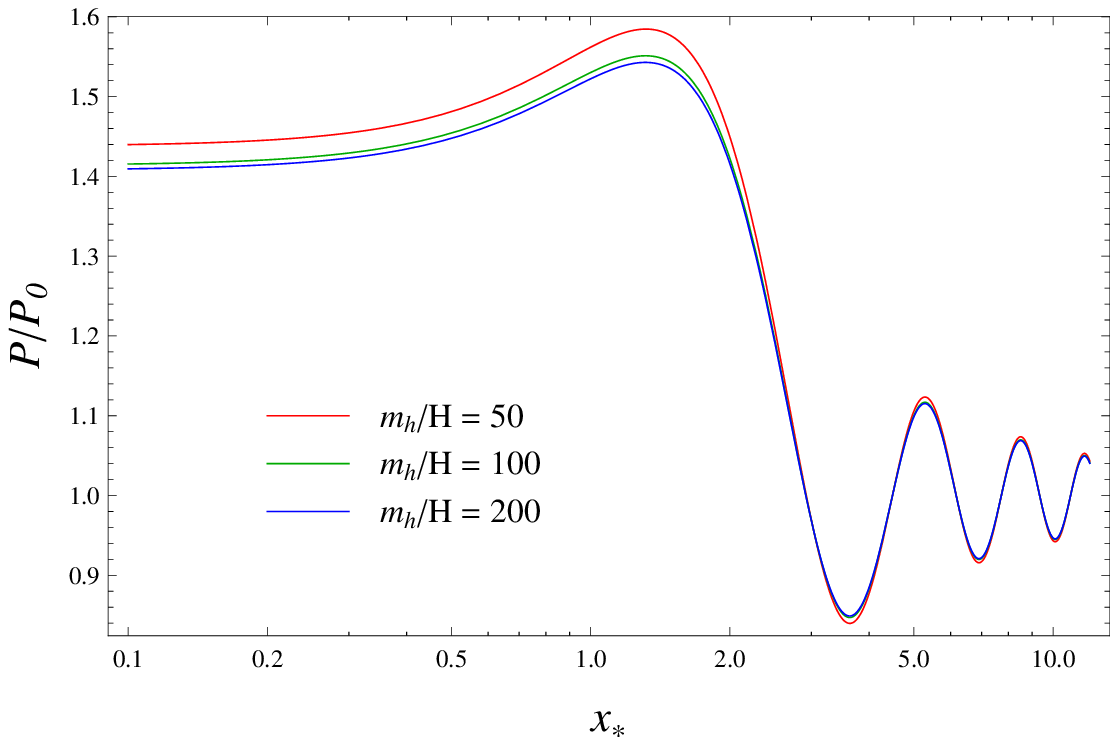}
    \end{minipage}
    $\qquad$
    \begin{minipage}{0.46\textwidth}
    \centering
        \includegraphics[width=\textwidth]{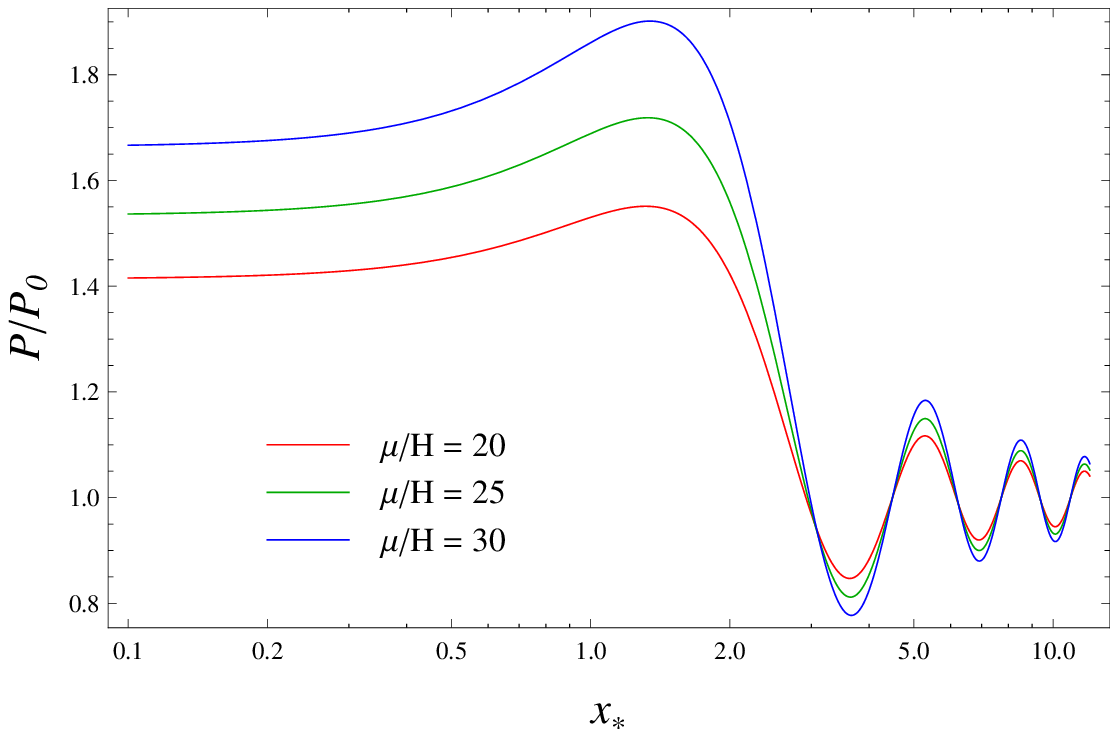}
    \end{minipage}
\caption{Power spectrum  (\ref{D_eff}) obtained from the effective theory. The plots of the left panel correspond to  different values of  $m_h/H$ (with fixed parameters $\tmu=20$ and $\Delta \theta = \pi/10$). The plots on the right panel correspond to  different values  of $\tmu$ (with fixed parameters $m_h/H=100$, $\Delta \theta = \pi/10$).}
            \label{fig:Dkeff}
\end{minipage}
\end{figure}
For modes that exited the Hubble radius before the turn, i.e. $x_*\ll 1$, the power spectrum is almost constant. For large values of $x_*$, corresponding to modes that exited the Hubble radius after the turn, the power spectrum oscillates around its unperturbed value, with the  amplitude of these oscillations  suppressed for large values of $x_*$. The power spectrum reaches its maximum value around $x_*\approx 1$, i.e. for modes that crossed the Hubble radius during the turn.

One also observes on the left panel that the power spectrum slightly changes with the value of $m_h/H$, in agreement with the above remark that the massive mode contribution is subdominant. An increase of $m_h/H$ corresponds to a slight decrease of the amplitude.  By contrast, the power spectrum is much more sensitive on the parameter $\tmu$, as illustrated on the right panel.
This means that the enhancement of the power spectrum is stronger for sharper turns, as expected.

While we have focussed our discussion on sharp turns, i.e. $\tmu\gg 1$, it is worth mentioning that, in the opposite limit $\tmu \ll 1$,  the turn lasts for several e-folds and  one recovers  the constant-coupling  case, discussed e.g. in \cite{Achucarro:2010jv}. In our notation, this corresponds to a constant $\dot{\theta}_m \approx \Delta\theta\,  {\mu}/{\sqrt{2\pi}}$,
    leading to
    \begin{eqnarray}
    c_s^2 \approx 1-\frac{2(\Delta\theta)^2 \mu^2}{\pi m_h^2},\qquad m_{\mathrm{eff}}^{2} &\approx & -\frac{(\Delta\theta)^2 \mu^2}{2\pi}\left( 1- \frac{4H^2}{m_h^2}\right).
    \end{eqnarray}
The resulting theory is analogous to  a single field theory with slightly modified sound speed and spectrum tilt.

Let us finally compare the results obtained from the effective theory with those derived from the two-field analysis.
A crucial remark is that, in  the decoupling limit  $m_h\rightarrow \infty$, the heavy mode contribution in (\ref{Dk_case2_app}) becomes negligible  and the result coincides with the dominant contribution (\ref{Dk_eff_0}) of the effective theory. 
This confirms the expectation that when the turn is not too sharp, the effective theory is sufficient to describe the power spectrum. When the sharpness of the turn reaches a critical value, the effective theory breaks down and an additional contribution appears in the power spectrum. 

\subsection{Comparison with the concrete model}
We now compare our semi-analytical exploration of the power spectrum with the fully numerical results for the explicit potential discussed in \ref{sec:model_num}. Here, in order to compare with the results in the previous
subsection, we set the following parameter values,
$M =6.0  \times 10^{-4}$,  $\Delta \theta = \pi/30$, 
while the other parameters are unchanged with respect to  the calculations of Section~\ref{sec:model_num}.

By solving simultaneously  the background equations and the perturbation equations, one can compute numerically the evolution of the perturbations. The  initial time is chosen well before the beginning of the turn, when the mode with the largest wavelength is well inside the heavy mass horizon. And the final time is chosen well after the turn, several e-folds after the mode with the smallest wavelength has exited the Hubble radius. At this final time, one can evaluate the power spectrum $\mathcal{P}(k)$ by adding the two  independent contributions obtained from respectively light or heavy initial conditions for the perturbations.

\begin{figure}[h]
\begin{minipage}{0.98\textwidth}
    \centering
  \begin{minipage}{0.4\textwidth}
    \centering
        \includegraphics[width=\textwidth]{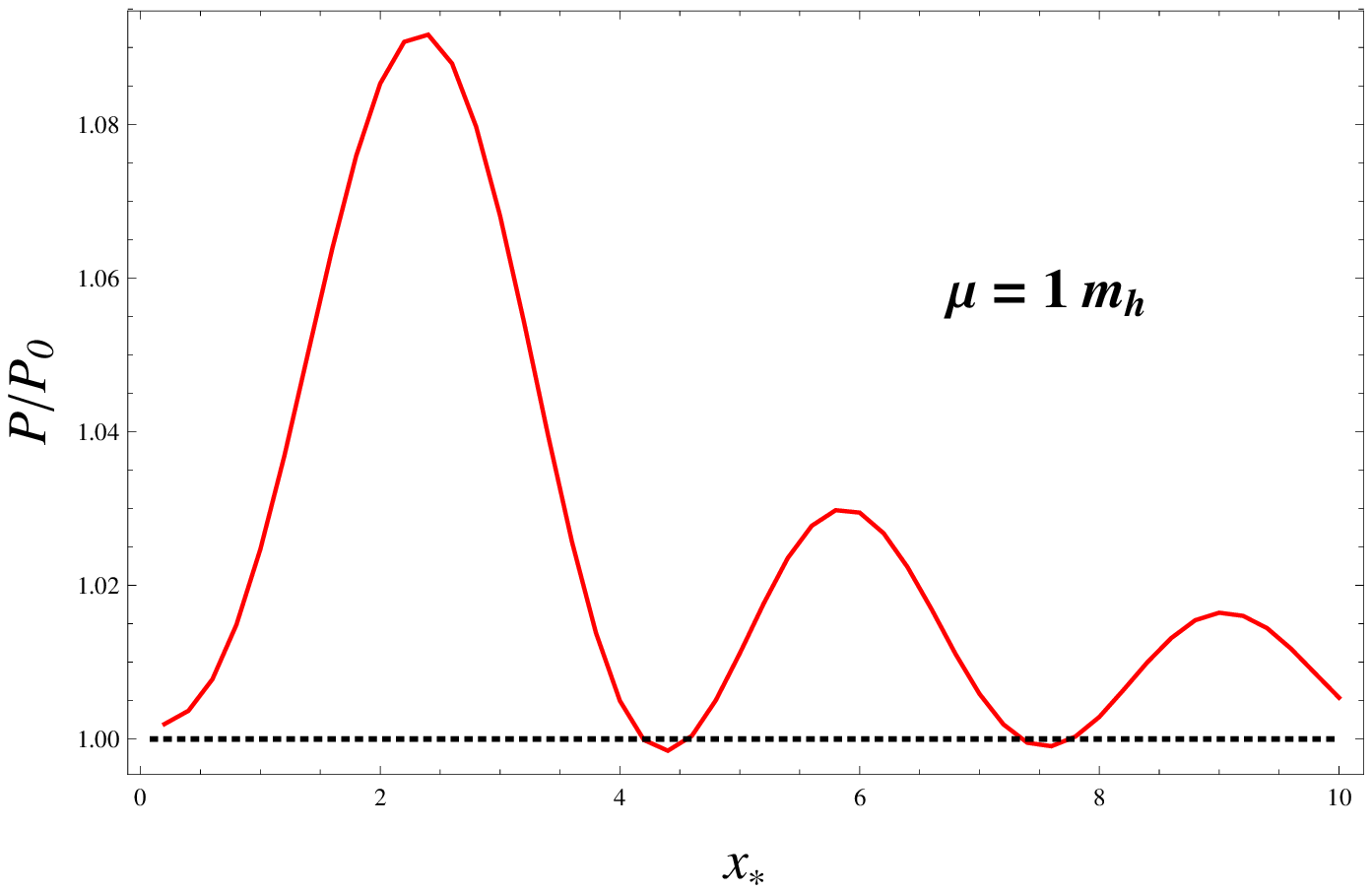}
        \end{minipage}
    $\qquad$
    \begin{minipage}{0.4\textwidth}
    \centering
        \includegraphics[width=\textwidth]{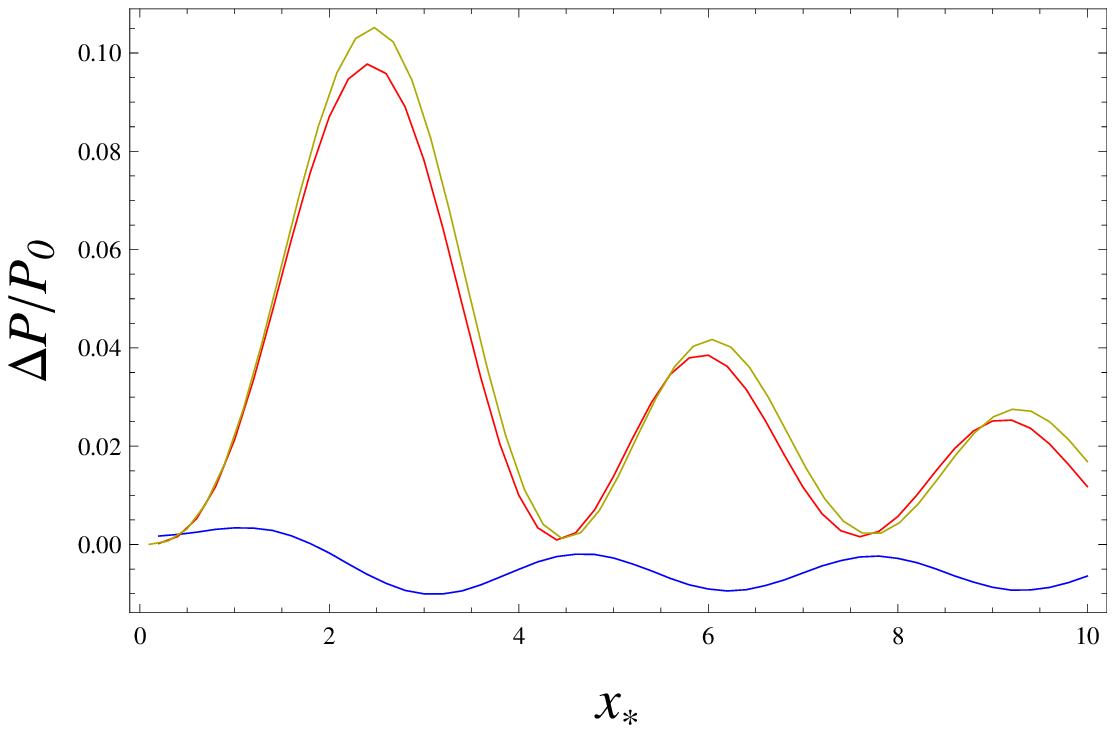}
    \end{minipage}
       \begin{minipage}{0.4\textwidth}
    \centering
        \includegraphics[width=\textwidth]{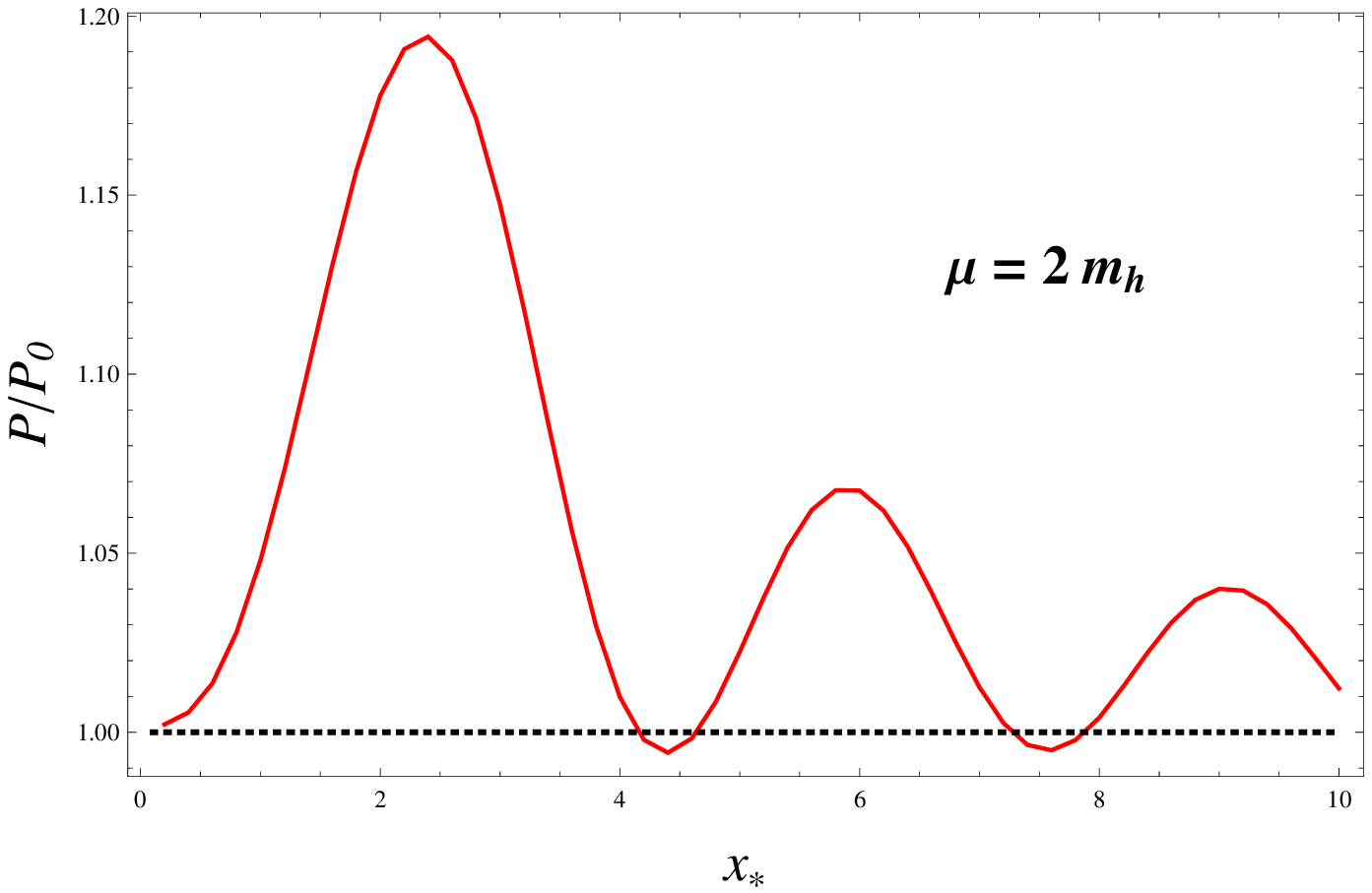}
                \end{minipage}
    $\qquad$
    \begin{minipage}{0.4\textwidth}
    \centering
 \includegraphics[width=\textwidth]{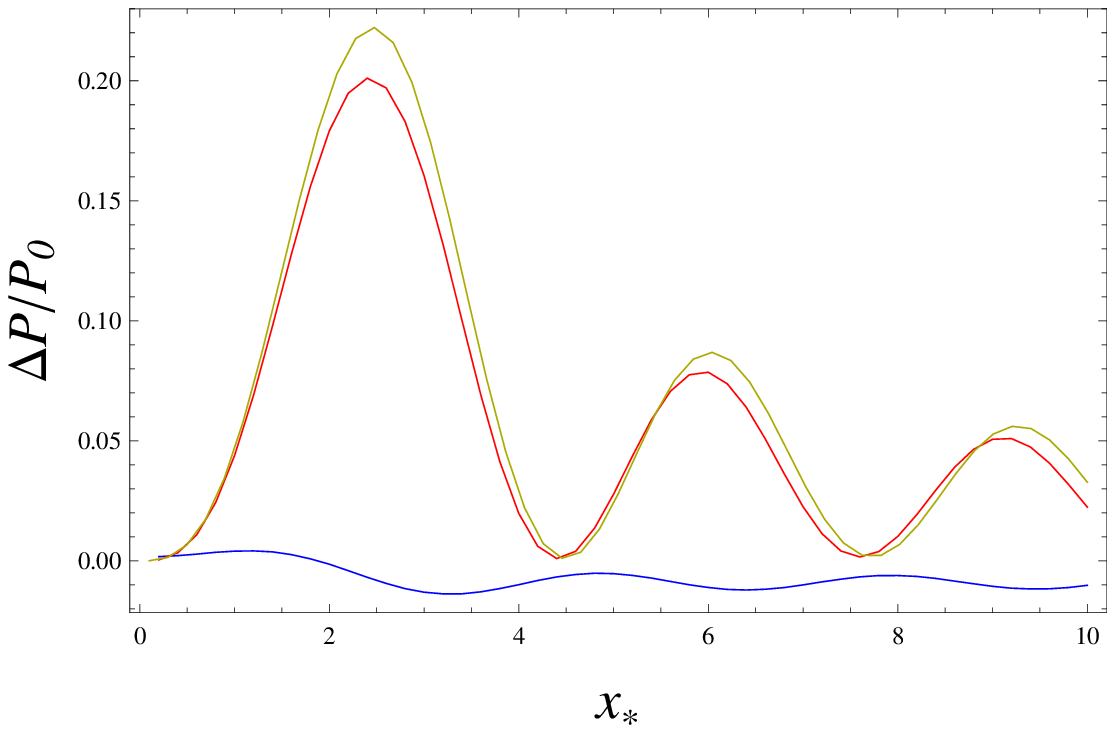}
    \end{minipage}
            \caption{Power spectrum from the concrete model. The total power spectrum is plotted on the left panel. On the right panel, one finds the light contribution (blue), the heavy contribution (red) and, for comparison, our analytical expression for the heavy contribution (yellow). From top to bottom: $s=3000 \sqrt{3}$ and $6000 \sqrt{3}$
which correspond respectively to $\mu = m_h $ and $\mu = 2 m_h$.}
            \label{fig:spectrum_concrete}
\end{minipage}
\end{figure}
In Fig.~\ref{fig:spectrum_concrete} we have plotted the power spectrum for two different values of the parameter $s$, which correspond approximately to the values 
$\mu=m_h$ and $\mu=2 m_h$.  
In the right panels, we show separately the light and heavy contributions to the total power spectrum, as well as our analytical approximation for the heavy contribution. One can observe that the analytical approximation gives a good estimate of the heavy contribution. We have not plotted the light contribution, which is significantly different from our analytical approximation. This difference is not surprising since the light mass $m_l$ is assumed to be zero in our simplified analysis whereas it fluctuates during the turn in our concrete example, and the final light contribution  is very sensitive to $m_l$.

\section{Conclusion}

In this work, we have studied the impact on inflation of heavy modes, which are expected to be ubiquitous in the context of high energy physics models of inflation that lead to a multi-field landscape.
 A crucial ingredient of  our investigation is the use of the mass basis, which consists of the eigenvectors of the effective mass matrix for the linear perturbations, instead of the traditional kinematic basis.
 Within this basis, we have derived the {\it multi-dimensional} low energy effective theory governing the light degrees of freedom, starting from an initial system with an arbitrary number of light modes and of heavy modes.  Our result thus generalizes the results of \cite{Achucarro:2010jv,Achucarro:2010da}, where a single light field was assumed.

We have then explored the imprints of these massive modes when the inflationary trajectory undergoes a sudden turn. Assuming a simple Gaussian ansatz for the time evolution of the angular velocity, parametrized by the mass scale $\mu\sim 1/\Delta t_{\rm turn}$, we have been able to obtain an analytical description of the evolution of the velocity direction with respect to the lightest direction of the potential.  The sharpness of the turn is governed by the ratio $\mu/m_h$. When this ratio is of order one or higher, the turn is sharp and the angle between the velocity and the light direction undergoes oscillations, with frequency $\omega\sim m_h$ and damped by expansion ($\propto a^{-3/2}$),  during the relaxation phase after the turn. We have confirmed this behavior by numerical integration for a concrete model.

We have also investigated in detail how the turn affects the final adiabatic power spectrum.    We have confirmed that, for a soft turn,  the low energy effective description is sufficient to account for the power spectrum. However, for a sharp turn, the power spectrum contains several  contributions, which we have been able to estimate  analytically or numerically in our simplified description.The resulting power spectrum exhibits oscillations periodic in $k$ and it would 
interesting to  look for this type of signature in the CMB data, following previous investigations of oscillatory features (see e.g. \cite{Martin:2004yi,Ichiki:2009xs,Aich:2011qv,Meerburg:2011gd}{\footnote{Note that   the reference \cite{Meerburg:2011gd}
repeats the analysis introduced  in \cite{Flauger:2009ab}, in the context of monodromy inflation, with  WMAP7 data.}}).

Another aspect, which we have not considered  here, is the impact  of the tiny oscillatory component of the {\it background} due to the  oscillating trajectory in case of a sharp turn. This effect, recently discussed in \cite{Chen:2011zf,Chen:2011tu,Chen:2012ja}, is complementary to our analysis. It would be interesting to estimate  this effect with our formalism, taking into account explicitly  the coupling between the light and heavy modes at the level of linear perturbations.

Several extensions of the present work  can be envisaged in the future. An immediate generalization would  be to extend our results to non-standard kinetic terms, by introducing a non-trivial field space metric.  It would be also interesting to refine our  description of the turn by allowing additional light or heavy directions or by going beyond our simple ansatz for the angular velocity, in order to explore whether some features of  the power spectrum are sensitive to the details of the turn. Finally, it would be worth going beyond the linear perturbations to investigate the specific imprints of heavy modes on non-Gaussianities.


\acknowledgments
We would like to thank Krzysztof Turzynski for a crucial remark, as well as  
Xingang Chen and S\'ebastien Renaux-Petel for very interesting discussions.
X.G. and D.L. were partly supported by ANR (Agence Nationale de la Recherche) grant ``STR-COSMO" ANR-09-BLAN-0157-01. S.M. was supported by Labex P2IO in Orsay.

\appendix

\section{Analytical approximations for the power spectrum}{\label{sec:pert_app_detail}}

In this section, we give some details about the derivation of  the analytical approximations given in the main body of the paper in Sec.\ref{sec:perts}. The goal is to solve analytically the integrals $I_l$ and $I_h$, defined in (\ref{F_l}) and (\ref{F_h}), by using  the ansatz
\beq
\dot{\theta}_m(t) = \Delta\theta \frac{\mu}{\sqrt{2\pi}} e^{-\frac{1}{2}\mu^2t^2}\,.
\eeq
The correspondence between the cosmic time $t$ and the conformal time $\eta$ is given by
\beq
a=-\frac{1}{H\eta}=e^{Ht}\quad \implies\quad \eta=\eta_* e^{-Ht}
\eeq
where we fix $t_*=0$ for simplicity.
As a consequence,  we can express $\theta'_m$ and $\theta_m''$ as Gaussian  functions of the cosmic time $t$:
    \begin{eqnarray}
\theta_{m}'(t) & = & a_{\ast}\Delta\theta\frac{\mu}{\sqrt{2\pi}}e^{-\frac{1}{2}\mu^{2}t^{2}+Ht},\label{thetam_prime_t}\\
\theta_{m}''(t) & = & \frac{\Delta\theta}{\sqrt{2\pi}}a_{\ast}^{2}\mu H\left(1-\frac{\mu^{2}}{H}t\right)e^{-\frac{1}{2}\mu^{2}t^{2}+2Ht}.\label{thetam_prime2_t}
\end{eqnarray}
where $a_{\ast}\equiv-(H\eta_{\ast})^{-1}$ is the scale factor at the turning point.

\subsection{Sub-horizons scales}
Deep inside the two horizons, the mode functions are approximated by
\beq
u_{l(0)} \approx u_{h(0)}\approx \frac{e^{-ik\eta}}{\sqrt{2k}}=\frac{e^{ix_{\ast}e^{-Ht}}}{\sqrt{2k}}\,.
\eeq
Since the integrands in $I_l$ and $I_h$ are exponentially suppressed  when $|t|>\mu^{-1}$, it is appropriate to use the approximation $e^{-Ht} \approx 1- Ht+\frac{1}{2}H^2t^2$ when $H\ll \mu$, which  will be assumed here.
We thus substitute in the integrals
    \begin{equation}{\label{ul_uh_app_case1}}
        u_{l(0)}\approx u_{h(0)}\approx\frac{e^{ix_{\ast}e^{-Ht}}}{\sqrt{2k}}\approx\frac{e^{ix_{\ast}\left(1-Ht+\frac{1}{2}H^{2}t^{2}\right)}}{\sqrt{2k}},
    \end{equation}
as well as
    \begin{equation}{\label{uh_prime_app_case1}}
        u_{h(0)}' \approx a_{\ast}\frac{ix_{\ast}}{\sqrt{2k}}e^{ix_{\ast}\left(1-Ht+\frac{1}{2}H^{2}t^{2}\right)+Ht}\left(-H+H^{2}t\right).
    \end{equation}

The corresponding integrals  $I_l$ and $I_h$, defined in (\ref{F_l}) and (\ref{F_h}), can then be evaluated analytically. One finds
    \begin{eqnarray}
I_{l}(k) =  i\left(\Delta\theta\right)^{2}\frac{\tilde{\mu}}{4\sqrt{\pi}x_{*}}\left(e^{\frac{1}{4\tilde{\mu}^{2}}}+\left(1-i\frac{x_{*}}{\tilde{\mu}^{2}}\right)^{-\frac{1}{2}}e^{\frac{4x_{*}^{2}+\tilde{\mu}^{2}\left(1-4x_{*}^{2}\right)}{4\left(\tilde{\mu}^{4}+x_{*}^{2}\right)}+i\frac{x_{*}\left(1-4\tilde{\mu}^{2}+8\tilde{\mu}^{4}+4x_{*}^{2}\right)}{4\left(\tilde{\mu}^{4}+x_{*}^{2}\right)}}\right),\label{Il_app_case1_ori}
\end{eqnarray}
and
\begin{equation}
I_{h}(k)=\Delta\theta \, e^{\frac{1}{2\tilde{\mu}^{2}}}\left(1-\tilde{\mu}^{-2}\right).\label{Ih_app_case1_ori}
\end{equation}

\subsection{Intermediate scales ($1\lesssim x_* \lesssim \nu$)}
For the light mode, we start from  the exact form of the mode function (\ref{ul0}) and  use the same approximation for the exponential factor as in the previous subsection, which gives
    \begin{equation}{\label{ul0_app_case2}}
        u_{l(0)}=\frac{e^{-ik\eta}}{\sqrt{2k}}\left(1-\frac{i}{k\eta}\right)\approx\frac{e^{ix_{\ast}\left(1-Ht+\frac{1}{2}H^{2}t^{2}\right)}}{\sqrt{2k}}\left(1+\frac{i}{x_{\ast}}e^{Ht}\right)\,.
    \end{equation}
For the heavy mode, we use the approximation (\ref{uh_app}), which yields
    \begin{eqnarray}
u_{h(0)} & \approx & \frac{A}{\sqrt{2k}}e^{i\nu\ln\left(-k\eta\right)}\sqrt{-k\eta}=\frac{A}{\sqrt{2k}}\sqrt{x_{\ast}}e^{i\nu\ln x_{\ast}}e^{-\left(\frac{1}{2}+i\nu\right)Ht},\qquad A\equiv\frac{\sqrt{2\pi}e^{-\frac{\pi}{2}\nu}}{\Gamma(i\nu+1)}e^{i\left(\frac{\pi}{4}-\nu\ln2\right)} \label{uh0_app_case2}\\
u_{h(0)}' & = & a\frac{du_{h(0)}}{dt}\approx-a_{\ast}\left(\frac{1}{2}+i\nu\right)H\frac{A}{\sqrt{2k}}\sqrt{x_{\ast}}e^{i\nu\ln x_{\ast}}e^{\left(\frac{1}{2}-i\nu\right)Ht}\,.\label{uh0_prime_app_case2}
\end{eqnarray}
One finds that the light integral can be written in the form
  \begin{equation}{\label{Il_case2_ori}}
        I_{l}\left(k\right)=\frac{i\left(\Delta\theta\right)^{2}\tilde{\mu}}{4\sqrt{\pi}x_{*}^{3}}
        \left[\left(1-i\frac{x_{*}}{\tilde{\mu}^{2}}\right)^{-1/2}\left(e^{p_{1}+iq_{1}}x_{*}^{2}+2ie^{p_{2}+iq_{2}}x_{*}-e^{p_{3}+iq_{3}}\right) +e^{\frac{1}{4\tilde{\mu}^{2}}}\left(e^{\frac{2}{\tilde{\mu}^{2}}}+x_{*}^{2}\right)\right],
    \end{equation}
with
    \begin{eqnarray}
p_{1} & = & \frac{\tilde{\mu}^{2}-4\left(\tilde{\mu}^{2}-1\right)x_{*}^{2}}{4\left(\tilde{\mu}^{4}+x_{*}^{2}\right)},\qquad p_{2}=\frac{\tilde{\mu}^{2}-\left(\tilde{\mu}^{2}-2\right)x_{*}^{2}}{\tilde{\mu}^{4}+x_{*}^{2}},\qquad p_{3}=\frac{9\tilde{\mu}^{2}-4\left(\tilde{\mu}^{2}-3\right)x_{*}^{2}}{4\left(\tilde{\mu}^{4}+x_{*}^{2}\right)},\label{p123}\\
q_{1} & = & \frac{x_{*}\left(1-4\tilde{\mu}^{2}+8\tilde{\mu}^{4}+4x_{*}^{2}\right)}{4\left(\tilde{\mu}^{4}+x_{*}^{2}\right)},\qquad q_{2}=\frac{x_{*}\left(1-2\tilde{\mu}^{2}+2\tilde{\mu}^{4}+x_{*}^{2}\right)}{\tilde{\mu}^{4}+x_{*}^{2}},\qquad q_{3}=\frac{x_{*}\left(9-12\tilde{\mu}^{2}+8\tilde{\mu}^{4}+4x_{*}^{2}\right)}{4\left(\tilde{\mu}^{4}+x_{*}^{2}\right)}.\label{q123}
\end{eqnarray}
And the heavy contribution is given by
    \begin{equation}{\label{Ih_case2_ori}}
        I_{h}\left(k\right)=\frac{\Delta\theta}{4\sqrt{\nu}x_{*}^{3/2}}e^{i\nu(1-\ln\left(2\nu\right))}e^{i\nu\ln x_{\ast}}\left[\left(1+i\frac{x_{*}}{\tilde{\mu}^{2}}\right)^{-3/2}f_{+}(x_{\ast})
        + \left(1-i\frac{x_{*}}{\tilde{\mu}^{2}}\right)^{-3/2} f_{-}(x_{\ast}) \right],
    \end{equation}
with
    \begin{eqnarray}
f_{+}\left(x_{\ast}\right) & = & e^{p_{1+}+iq_{1+}}\left(4\frac{\nu}{\tilde{\mu}^{2}}x_{*}-3-2i\left(\nu+x_{*}\right)\right)+ie^{p_{2+}+iq_{2+}}x_{*}\left(4\frac{\nu}{\tilde{\mu}^{2}}x_{*}-1-2i\left(\nu+x_{*}\right)\right),\label{f_plus_def}\\
f_{-}\left(x_{\ast}\right) & = & e^{p_{1-}+iq_{1-}}\left(4\frac{\nu}{\tilde{\mu}^{2}}x_{*}+3+2i\left(\nu-x_{*}\right)\right)-ie^{p_{2-}+iq_{2-}}x_{*}\left(4\frac{\nu}{\tilde{\mu}^{2}}x_{*}+1+2i\left(\nu-x_{*}\right)\right),\label{f_minus_def}
\end{eqnarray}
where
    \begin{eqnarray}
p_{1+} & = & \frac{9\tilde{\mu}^{2}-4\left(x_{*}-\nu\right)\left[\left(\tilde{\mu}^{2}-3\right)x_{*}-\nu\tilde{\mu}^{2}\right]}{8\left(\tilde{\mu}^{4}+x_{*}^{2}\right)},\label{p1_plus}\\
p_{2+} & = & \frac{\tilde{\mu}^{2}-4\left(x_{*}-\nu\right)\left[\left(\tilde{\mu}^{2}-1\right)x_{*}-\nu\tilde{\mu}^{2}\right]}{8\left(\tilde{\mu}^{4}+x_{*}^{2}\right)},\label{p2_plus}\\
q_{1+} & = & \frac{-12\nu\tilde{\mu}^{2}+\left(-9+4\nu^{2}+12\tilde{\mu}^{2}-8\tilde{\mu}^{4}\right)x_{*}-8\nu x_{*}^{2}-4x_{*}^{3}}{8\left(\tilde{\mu}^{4}+x_{*}^{2}\right)},\label{q1_plus}\\
q_{2+} & = & \frac{-4\nu\tilde{\mu}^{2}+\left(-1+4\nu^{2}+4\tilde{\mu}^{2}-8\tilde{\mu}^{4}\right)x_{*}-8\nu x_{*}^{2}-4x_{*}^{3}}{8\left(\tilde{\mu}^{4}+x_{*}^{2}\right)},\label{q2_plus}
\end{eqnarray}
and
    \begin{eqnarray}
p_{1-} & = & \frac{9\tilde{\mu}^{2}-4\left(\nu+x_{*}\right)\left(\nu\tilde{\mu}^{2}+\left(\tilde{\mu}^{2}-3\right)x_{*}\right)}{8\left(\tilde{\mu}^{4}+x_{*}^{2}\right)},\label{p1_minus}\\
p_{2-} & = & \frac{\tilde{\mu}^{2}-4\left(\nu+x_{*}\right)\left(\nu\tilde{\mu}^{2}+\left(\tilde{\mu}^{2}-1\right)x_{*}\right)}{8\left(\tilde{\mu}^{4}+x_{*}^{2}\right)},\label{p2_minux}\\
q_{1-} & = & \frac{-12\nu\tilde{\mu}^{2}+\left(9-4\nu^{2}-12\tilde{\mu}^{2}+8\tilde{\mu}^{4}\right)x_{*}-8\nu x_{*}^{2}+4x_{*}^{3}}{8\left(\tilde{\mu}^{4}+x_{*}^{2}\right)},\label{q1_minus}\\
q_{2-} & = & \frac{-4\nu\tilde{\mu}^{2}+\left(1-4\nu^{2}-4\tilde{\mu}^{2}+8\tilde{\mu}^{4}\right)x_{*}-8\nu x_{*}^{2}+4x_{*}^{3}}{8\left(\tilde{\mu}^{4}+x_{*}^{2}\right)}.\label{q2_minus}
\end{eqnarray}
In the limit  $\tilde{\mu} \gg x_{\ast}$, (\ref{Il_case2_ori}) and (\ref{Ih_case2_ori}) simplify into  (\ref{I_l_app_case2}) and (\ref{I_h_app_case2}).

\subsection{Super-horizon scales ($x_*\ll 1$)}
In this case, (\ref{uh_app}) and thus (\ref{uh0_app_case2}) and (\ref{uh0_prime_app_case2}) remain  good approximations for the heavy mode. For the light mode,  it is convenient to expand the expression  for $u_{l(0)}$ with respect to  the  small parameter $x_{\ast}$ so that
    \begin{equation}
        u_{l(0)}=\frac{e^{-ik\eta}}{\sqrt{2k}}\left(1-\frac{i}{k\eta}\right)=\frac{e^{ix_{\ast}e^{-Ht}}}{\sqrt{2k}}\left(1+\frac{i}{x_{\ast}}e^{Ht}\right)\approx\frac{1}{\sqrt{2k}}\left(\frac{i}{x_{\ast}}e^{Ht}-\frac{1}{3}e^{-2Ht}x_{\ast}^{2}\right).
    \end{equation}
After some manipulations, we get
    \begin{eqnarray}
I_{l} & = & \left(\Delta\theta\right)^{2}\frac{\tilde{\mu}}{18\sqrt{\pi}}\left(3+ie^{\frac{9}{4\tmu^{2}}}x_{*}^{3}\right),\\
I_{h} & = & -\Delta\theta\frac{(2\nu+3i)}{6\sqrt{\nu}}e^{\frac{9-4\nu^2}{8\tilde{\mu}^{2}}+i\nu\left(1+\frac{3}{2\tilde{\mu}^{2}}\right)}x_{*}^{{3/2}}e^{i\nu\ln\left(\frac{x_{\ast}}{2\nu}\right)}\,.
\end{eqnarray}

\subsection{Effective theory}{\label{see:pert_eff_detail}}
 The initial condition for the  mode function $v_{(0)}$ is the same as in (\ref{ul0_app_case2}), namely
    \begin{equation}{\label{v_eff_app_fin}}
      v_{(0)}\left(t,k\right) \approx \frac{e^{ix_{\ast}\left(1-Ht+\frac{1}{2}H^{2}t^{2}\right)}}{\sqrt{2k}}\left(1+\frac{i}{x_{\ast}}e^{Ht}\right).
    \end{equation}
Substituting in  the integral $I(k)$ defined in (\ref{I_eff_2f}), and using (\ref{thetam_prime_t}-\ref{thetam_prime2_t})
with the expansions (\ref{cs_eff_exp}) and (\ref{meff_eff_exp}), we find, at leading order,
\beq
I_{0}\left(k\right)=\frac{\left(\Delta\theta\right)^{2}\tilde{\mu}}{4\sqrt{\pi}x_{*}^{3}}\left[i\left(e^{\frac{9}{4\tilde{\mu}^{2}}}+x_{*}^{2}e^{\frac{1}{4\tilde{\mu}^{2}}}\right)+\frac{\tilde{\mu}}{\sqrt{\tilde{\mu}^{2}-ix_{*}}}\left(-ie^{\frac{9i+4\left(3-2\tilde{\mu}^{2}+ix_{*}\right)x_{*}}{4\left(i\tilde{\mu}^{2}+x_{*}\right)}}-2x_{*}e^{\frac{i+\left(2-2\tilde{\mu}^{2}+ix_{*}\right)x_{*}}{i\tilde{\mu}^{2}+x_{*}}}+ix_{*}^{2}e^{\frac{i+4x_{*}-8\tilde{\mu}^{2}x_{*}+4ix_{*}^{2}}{4i\tilde{\mu}^{2}+4x_{*}}}\right)\right].
\eeq
and, at next-to-leading order,
 \begin{eqnarray}
I_{1} & = & \frac{H^{2}(\Delta\theta)^{4}\tilde{\mu}^{3}}{4\pi^{3/2}m_{h}^{2}\sqrt{2\tilde{\mu}^{2}-ix_{*}}x_{*}^{3}}e^{-\frac{4\tilde{\mu}^{2}x_{*}}{2i\tilde{\mu}^{2}+x_{*}}}\cr
 &  & \times\left[2e^{\frac{1-4ix_{*}+4x_{*}^{2}}{8\tilde{\mu}^{2}-4ix_{*}}}\tilde{\mu}\left(ie^{\frac{2-2ix_{*}}{2\tilde{\mu}^{2}-ix_{*}}}+2e^{\frac{3i+4x_{*}}{4\left(2i\tilde{\mu}^{2}+x_{*}\right)}}x_{*}-ix_{*}^{2}\right)-ie^{\frac{1}{8\tilde{\mu}^{2}}+\frac{4\tilde{\mu}^{2}x_{*}}{2i\tilde{\mu}^{2}+x_{*}}}\sqrt{4\tilde{\mu}^{2}-2ix_{*}}\left(e^{\frac{1}{\tilde{\mu}^{2}}}+x_{*}^{2}\right)\right]\cr
 &  & +\frac{H^{2}(\Delta\theta)^{2}\tilde{\mu}}{16\sqrt{\pi}m_{h}^{2}\left(\tilde{\mu}^{2}-ix_{*}\right){}^{5/2}x_{*}^{3}}e^{-\frac{x_{*}+2\tilde{\mu}^{2}x_{*}}{i\tilde{\mu}^{2}+x_{*}}}\cr
 &  & \times\Big\{ ie^{\frac{9}{4\tilde{\mu}^{2}}+\frac{x_{*}}{i\tilde{\mu}^{2}+x_{*}}+\frac{2\tilde{\mu}^{2}x_{*}}{i\tilde{\mu}^{2}+x_{*}}}\left(9+2\tilde{\mu}^{2}\right)\left(\tilde{\mu}^{2}-ix_{*}\right){}^{5/2}+32e^{\frac{\left(1+ix_{*}\right)x_{*}}{i\tilde{\mu}^{2}+x_{*}}}\tilde{\mu}x_{*}^{3}\left(i\tilde{\mu}^{2}+x_{*}\right){}^{2}\cr
 &  & -16ie^{\frac{1+4x_{*}^{2}}{4\tilde{\mu}^{2}-4ix_{*}}}\tilde{\mu}x_{*}^{4}\left(i\tilde{\mu}^{2}+x_{*}\right){}^{2}-ie^{\frac{x_{*}+8\tilde{\mu}^{4}x_{*}+\tilde{\mu}^{2}\left(i+4x_{*}\right)}{4\tilde{\mu}^{2}\left(i\tilde{\mu}^{2}+x_{*}\right)}}\sqrt{\tilde{\mu}^{2}-ix_{*}}x_{*}^{2}\left(i\tilde{\mu}^{2}+x_{*}\right){}^{2}\left(-15+2\tilde{\mu}^{2}+16x_{*}^{2}\right)\cr
 &  & +e^{\frac{1-8ix_{*}+4x_{*}^{2}}{4\tilde{\mu}^{2}-4ix_{*}}}\tilde{\mu}x_{*}^{2}\left(2i\tilde{\mu}^{6}+52ix_{*}^{2}+8i\tilde{\mu}^{2}x_{*}\left(13i+x_{*}\right)+\tilde{\mu}^{4}\left(-47i+14x_{*}-20ix_{*}^{2}\right)\right)\cr
 &  & +4e^{\frac{1-3ix_{*}+x_{*}^{2}}{\tilde{\mu}^{2}-ix_{*}}}\tilde{\mu}x_{*}\left(-\tilde{\mu}^{6}-18x_{*}^{2}-4\tilde{\mu}^{2}x_{*}\left(9i+x_{*}\right)+\tilde{\mu}^{4}\left(8+17ix_{*}+10x_{*}^{2}\right)\right)\cr
 &  & +ie^{\frac{9-16ix_{*}+4x_{*}^{2}}{4\tilde{\mu}^{2}-4ix_{*}}}\tilde{\mu}\left(-2\tilde{\mu}^{6}-36x_{*}^{2}-8\tilde{\mu}^{2}x_{*}\left(9i+x_{*}\right)+\tilde{\mu}^{4}\left(-9+54ix_{*}+20x_{*}^{2}\right)\right)\Big\}.
\end{eqnarray}




\begin{thebibliography}{99}

\bibitem{Tolley:2009fg}
  A.~J.~Tolley and M.~Wyman,
  Phys.\ Rev.\ D {\bf 81}, 043502 (2010)
  [arXiv:0910.1853 [hep-th]].


\bibitem{Achucarro:2010jv}
  A.~Achucarro, J.~O.~Gong, S.~Hardeman, G.~A.~Palma and S.~P.~Patil,
  Phys.\ Rev.\  D {\bf 84} (2011) 043502
  [arXiv:1005.3848 [hep-th]].


\bibitem{Achucarro:2010da}
  A.~Achucarro, J.~O.~Gong, S.~Hardeman, G.~A.~Palma and S.~P.~Patil,
  JCAP {\bf 1101} (2011) 030
  [arXiv:1010.3693 [hep-ph]].

\bibitem{Cremonini:2010ua}
  S.~Cremonini, Z.~Lalak and K.~Turzynski,
  JCAP {\bf 1103}, 016 (2011)
  [arXiv:1010.3021 [hep-th]].



\bibitem{Achucarro:2012sm}
  A.~Achucarro, J.~-O.~Gong, S.~Hardeman, G.~A.~Palma and S.~P.~Patil,
  arXiv:1201.6342 [hep-th].

\bibitem{Chen:2012ge}
  X.~Chen and Y.~Wang,
  arXiv:1205.0160 [hep-th].

\bibitem{Pi:2012gf}
  S.~Pi and M.~Sasaki,
  arXiv:1205.0161 [hep-th].


\bibitem{Avgoustidis:2012yc}
  A.~Avgoustidis, S.~Cremonini, A.~-C.~Davis, R.~H.~Ribeiro, K.~Turzynski and S.~Watson,
  arXiv:1203.0016 [hep-th].

\bibitem{Achucarro:2012yr}
  A.~Achucarro, V.~Atal, S.~Cespedes, J.~-O.~Gong, G.~A.~Palma and S.~P.~Patil,
  arXiv:1205.0710 [hep-th].

\bibitem{Shiu:2011qw}
  G.~Shiu and J.~Xu,
  Phys.\ Rev.\ D {\bf 84}, 103509 (2011)
  [arXiv:1108.0981 [hep-th]].

 
\bibitem{Cespedes:2012hu} 
  S.~Cespedes, V.~Atal and G.~A.~Palma,
  JCAP {\bf 1205}, 008 (2012)
  [arXiv:1201.4848 [hep-th]].

\bibitem{Jackson:2010cw}
  M.~G.~Jackson and K.~Schalm,
  Phys.\ Rev.\ Lett.\  {\bf 108} (2012) 111301
  [arXiv:1007.0185 [hep-th]].

\bibitem{Jackson:2011qg}
  M.~G.~Jackson and K.~Schalm,
  arXiv:1104.0887 [hep-th].

\bibitem{Jackson:2012fu}
  M.~G.~Jackson and K.~Schalm,
  arXiv:1202.0604 [hep-th].


\bibitem{Gordon:2000hv}
  C.~Gordon, D.~Wands, B.~A.~Bassett and R.~Maartens,
  Phys.\ Rev.\  D {\bf 63} (2001) 023506
  [arXiv:astro-ph/0009131].

\bibitem{GNvT}
  S.~Groot Nibbelink and B.~J.~W.~van Tent,
  Class.\ Quant.\ Grav.\ \ {\bf 19} (2002) 613
  [arXiv:hep-ph/0107272 [hep-ph]].



\bibitem{Chen:2009zp}
  X.~Chen and Y.~Wang,
  JCAP {\bf 1004}, 027 (2010)
  [arXiv:0911.3380 [hep-th]].


\bibitem{Gao:2009qy}
  X.~Gao,
  JCAP {\bf 1002} (2010) 019
  [arXiv:0908.4035 [hep-th]].

\bibitem{Chen:2009we}
  X.~Chen and Y.~Wang,
  Phys.\ Rev.\ D {\bf 81}, 063511 (2010)
  [arXiv:0909.0496 [astro-ph.CO]].



\bibitem{LRPST}
  D.~Langlois and S.~Renaux-Petel,
  JCAP\ {\bf 0804} (2008) 017
  [arXiv:0801.1085 [hep-th]].
  D.~Langlois, S.~Renaux-Petel, D.~A.~Steer and T.~Tanaka,
  Phys.\ Rev.\ Lett.\ \ {\bf 101} (2008) 061301
  [arXiv:0804.3139 [hep-th]].
  D.~Langlois, S.~Renaux-Petel, D.~A.~Steer and T.~Tanaka,
  Phys.\ Rev.\ D\ {\bf 78} (2008) 063523
  [arXiv:0806.0336 [hep-th]].
  F.~Arroja, S.~Mizuno and K.~Koyama,
  JCAP {\bf 0808}, 015 (2008)
  [arXiv:0806.0619 [astro-ph]].

\bibitem{Langlois:2010xc}
  D.~Langlois,
  Lect.\ Notes Phys.\  {\bf 800}, 1 (2010)
  [arXiv:1001.5259 [astro-ph.CO]].



\bibitem{Peterson:2011yt}
  C.~M.~Peterson and M.~Tegmark,
  arXiv:1111.0927 [astro-ph.CO].




\bibitem{LV}
  D.~Langlois and F.~Vernizzi,
  JCAP\ {\bf 0702} (2007) 017
  [astro-ph/0610064].


  \bibitem{AandS}
    M. Abramowitz and I. A. Stegun, eds. (1972), {\it Handbook of Mathematical Functions with Formulas, Graphs, and Mathematical Tables}, New York: Dover Publications, ISBN 978-0-486-61272-0.





\bibitem{Martin:2004yi}
  J.~Martin and C.~Ringeval,
  JCAP {\bf 0501}, 007 (2005)
  [hep-ph/0405249].

\bibitem{Ichiki:2009xs}
  K.~Ichiki, R.~Nagata and J.~'i.~Yokoyama,
  Phys.\ Rev.\ D {\bf 81} (2010) 083010
  [arXiv:0911.5108 [astro-ph.CO]].

\bibitem{Aich:2011qv}
  M.~Aich, D.~K.~Hazra, L.~Sriramkumar and T.~Souradeep,
  arXiv:1106.2798 [astro-ph.CO].

\bibitem{Meerburg:2011gd}
  P.~D.~Meerburg, R.~Wijers and J.~P.~van der Schaar,
  arXiv:1109.5264 [astro-ph.CO].


\bibitem{Flauger:2009ab}
  R.~Flauger, L.~McAllister, E.~Pajer, A.~Westphal and G.~Xu,
  JCAP {\bf 1006} (2010) 009
  [arXiv:0907.2916 [hep-th]].



\bibitem{Chen:2012ja}
  X.~Chen and C.~Ringeval,
  arXiv:1205.6085 [astro-ph.CO].



\bibitem{Chen:2011zf}
  X.~Chen,
  arXiv:1104.1323 [hep-th].

\bibitem{Chen:2011tu}
  X.~Chen,
  Phys.\ Lett.\ B {\bf 706}, 111 (2011)
  [arXiv:1106.1635 [astro-ph.CO]].



\end{thebibliography}
\end{document}